\documentclass{aa}
\usepackage{epsf,epsfig}

\def\kms{\,km\thinspace s$^{-1}$ }  

\begin{document}

\titlerunning{VLT surface photometry and isophotal analysis of early-type
dwarfs}

\title{VLT surface photometry and isophotal analysis of early-type dwarf
galaxies in the Virgo cluster
\thanks{Based on observations collected at the European Southern Observatory
(ESO 63.O-0055 and 65.N-0062)}}

\author{Fabio D. Barazza \inst{1}
\and Bruno Binggeli \inst{1}
\and Helmut Jerjen \inst{2}}

\offprints{F.D. Barazza, e-mail : barazza@astro.unibas.ch}

\institute{Astronomisches Institut, Universit\"at Basel, Venusstrasse 7,
CH-4102 Binningen, Switzerland
\and
Research School of Astronomy and Astrophysics, The Australian National
University, Mt Stromlo Observatory, Cotter Road, Weston ACT 2611, Australia}

\date{Received 29 January 2003 / Accepted 20 May 2003}

\abstract{We have carried out surface photometry and an isophotal analysis for
a sample of 25 early-type dwarf (dE and dS0) 
galaxies in the Virgo cluster based on CCD images 
taken at the VLT with FORS1 and FORS2. For each galaxy we present $B$ 
and $R$-band surface brightness profiles, as well as the radial colour ($B-R$)
profile. 
We give total apparent $BR$ magnitudes, effective radii, effective surface 
brightnesses and total colour indices. The light profiles have been fitted with 
S\'ersic models and the corresponding parameters are compared to the ones for 
other classes of objects. In general, dEs and dS0s bridge 
the gap in parameter space between the giant ellipticals and the low-luminosity 
dwarf 
spheroidals in the Local Group, in accordance with previous findings. 
However, the observed profiles of the brightest cluster dwarfs show 
significant deviations from a simple S\'ersic model, indicating that there is 
more inner structure than just a nucleus. This picture is reinforced by our
isophotal analysis where complex radial 
dependencies 
of ellipticity, position angle, and isophotal shape parameter $a_4$ 
are exhibited not only by objects like IC3328, for which the presence of 
a disk component has been confirmed, but by many apparently normal dEs as well. 
In addition, we find a relation between the effective surface 
brightness, at a given luminosity, and the strength of the offset of the 
galaxy's nucleus with respect 
to the center of the isophotes. Dwarfs with large nuclear offsets also tend 
to have stronger isophotal twists. However, such twists are 
preferentially found in apparently round ($\epsilon < 0.3$) galaxies
and are always accompanied by significant radial changes of the ellipticity, 
which clearly
points to a projection effect. In sum, our findings suggest
the presence of substructure in most, and preferentially in the less
compact, bright early-type dwarfs. 
The physical (dynamical) meaning of this has yet to be explored.
\keywords{
galaxies: fundamental parameters --
galaxies: photometry -- galaxies: structure
galaxies: clusters: individual: Virgo -- 
galaxies: dwarf --
galaxies: elliptical and lenticular -- 
}
}
\maketitle

\section{Introduction}
Dwarf elliptical galaxies (hereafter dEs, subsuming ``dwarf spheroidals'' and
dwarf S0s) are by far the most numerous type of galaxy in the local universe 
(see Ferguson \& Binggeli (1994) for a review). As they prefer the high density
environment of galaxy clusters, the most suitable places for their study are the 
two nearest clusters in Virgo and Fornax, where catalogues of dEs have been 
established as part of extensive surveys (Binggeli et al.~1985, hereafter VCC; 
Ferguson 1989). The numerical dominance of dEs in these two 
galaxy aggregates is evident in the studies of the corresponding
galaxy luminosity functions (Sandage et al.~1985, Ferguson \& Sandage 1988).
A host of photographic and CCD studies of Virgo and Fornax dwarfs
in the eighties and early nineties led to our knowledge of the
basic photometric properties of dwarf ellipticals (Caldwell 1983,
Binggeli et al.~1984, Bothun et al.~1986, Caldwell \& Bothun 1987, 
Binggeli \& Cameron 1991, 1993). Little dE photometry
has been added to this until recently. 
Ryden et al.~(1999), still working with low-resolution CCD images, went
a step further by analyzing the isophotal shapes of a large sample 
of dEs, finding many dwarfs to be ``disky'' and ''boxy'' just as the giants.  
Miller et al.~(1998) observed 24 Virgo dwarfs with the HST to derive the 
specific 
Globular Cluster Frequency ($S_N$) for dEs. Based on the same high-resolution 
HST
images, Stiavelli et al.~(2001) analyzed the innermost regions (cusp slopes) of
these galaxies. However, it proved difficult to connect the central properties
with the {\em global}\/ dwarf structures; the drawback of HST here is clearly 
the small field of view.

On the other hand, enormous progress has recently 
been made at the kinematic ``frontline''. After almost a decade of 
stagnation
of kinematic measurements (see Ferguson \& Binggeli 1994), several groups are
now reporting their (partially conflicting) results on the (non-) rotational 
properties of early-type dwarfs. 
For instance, Geha et al.~(2001) used Keck II to measure rotation 
profiles for six Virgo dwarfs. No evidence of significant rotation was 
found among the target objects. A similar programme is being run at the VLT
(de\,Rijcke et al.~2002). The earlier conjecture that
dwarf ellipticals in general are not rotation-supported is definitively 
confirmed
by these studies (also Thomas et al.~2003). However, it has also become 
clear that among the brightest cluster early-type dwarfs, in particular the 
dS0s,
there are many rotation-supported systems (Simien \& Prugniel 2002).  

There are also {\em photometric}\/ hints about the existence of disk 
galaxies
among the bright cluster early-type dwarfs. Jerjen, Kalnajs \& Binggeli (2000, 
2001) 
discovered weak spiral structures and a bar in two Virgo dEs 
and subsequently Barazza et al.~(2002) found spiral and bar features in three 
additional objects. These findings are in fact based on the same VLT images for
which surface photometry is presented in the present paper. Deep,
high-resolution VLT imaging, providing a sufficiently large field of view,
is certainly ideally suited for the study of
the photometric properties of Virgo and Fornax dEs. 
With the present surface photometry and isophotal analysis of 25 early-type 
Virgo 
dwarfs 
based on high-quality $B$ and $R$ VLT images we aim at a more
systematic exploration of the structural complexity of dwarf ellipticals.
We especially address the question whether a 
S\'ersic model is an appropriate representation of the empirical surface 
brightness 
profiles of dEs. The isophotal 
analysis 
is used to derive ellipticity, position angle and isophotal shape profiles. 
In addition, we map nuclear offsets and isophotal twists of the sample dEs.
Overall, our findings show that these seemingly dull stellar systems are quite 
complex in structure. Dwarf elliptical galaxies are 
neither the scaled down version of giant ellipticals nor simply the final
state of a star forming dwarf irregular galaxy that has converted all its gas 
into stars.

The plan of this paper is as follows. In Sect.~2 we introduce the dwarf galaxy 
sample 
and provide some global photometric parameters. The data reduction and 
photometric 
calibration are described in Sect. 3. Sect.~4 and 5 are dedicated to the surface 
photometry and the isophotal analysis. The discussion and summary is given in 
Sect. 6.

\section{Sample and observations}
The dE galaxies studied here were originally chosen from the Virgo Cluster 
Catalog (VCC; Binggeli et al.~1985) for the purpose of measuring their distances
by means of the Surface Brightness Fluctuation method to explore the 
3-dimensional structure of the Virgo cluster (Jerjen et al.~2003). As it is a 
main requirement for the successful application of this method, galaxies 
were primarily selected on their morphological appearance, i.e. type ``dE'' or 
``dS0'', and on their apparent size, i.e. an isophotal 
radius $r_{B, 25}>30''$. Dwarf S0s are a frequent morphological variation
of the dE class among bright dwarfs, being characterized by the 
presence of an S0-like two-component structure, of some
bar-like feature, strongly twisting isophotes, or simply by high flattening.
In their
mean surface brightness profile, dS0s are indistinguishable from dEs
(Sandage \& Binggeli 1984, Binggeli \& Cameron 1991). 

Within these constraints, the sample was selected so as to get a 
good coverage in velocity space 
($-730$\kms$ <V_\odot<$1850\kms) and in the 
celestial distribution ($12^h 09^m  
< \mbox{R.A.(2000)} < 12^h 32^m$; $+08^\circ 26' < \mbox{Decl.(2000)} < 
+15^\circ 45'$). The core sample contained 16 bright early-type dwarfs which, 
however, could be increased by 9 more dwarfs, as these happened to lie in 
the field of view of the CCD. Among these are five rather faint dwarfs
(VCC0850, VCC0962, VCC0998, VCC1093, VCC1129). The total sample considered 
here is comprised therefore of 25 objects, 22 of which were imaged in $B$ and
$R$ filters (or $R_s$, which is the corresponding filter used on FORS2; in the
following we only use $R$) and three (IC3303, IC3518, UGC7854) in
$R$ only.

\begin{table*}[t]
\caption[]{Basic data and model-free parameters of the early-type dwarfs
considered in this study (extinction-corrected)}
\vspace{0.3 cm}
\begin{center}
\begin{tabular}{llllrlllllllc}
\hline
\\
\multicolumn{1}{l}{VCC} &
\multicolumn{1}{l}{Name} &
\multicolumn{1}{l}{Type} &
\multicolumn{1}{c}{$M_{B_T}$} &
\multicolumn{1}{c}{$v_{\odot}$} &
\multicolumn{1}{c}{$B_T$} &
\multicolumn{1}{c}{$r_{eff}^B$} &
\multicolumn{1}{c}{$\mu_{eff}^B$} &
\multicolumn{1}{c}{$R_T$} &
\multicolumn{1}{c}{$r_{eff}^R$} &
\multicolumn{1}{c}{$\mu_{eff}^R$} &
\multicolumn{1}{c}{$B-R$} &
\multicolumn{1}{c}{Notes} \\
\multicolumn{1}{l}{(1)} &
\multicolumn{1}{l}{(2)} &
\multicolumn{1}{l}{(3)} &
\multicolumn{1}{c}{(4)} &
\multicolumn{1}{c}{(5)} &
\multicolumn{1}{c}{(6)} &
\multicolumn{1}{c}{(7)} &
\multicolumn{1}{c}{(8)} &
\multicolumn{1}{c}{(9)} &
\multicolumn{1}{c}{(10)} &
\multicolumn{1}{c}{(11)} &
\multicolumn{1}{c}{(12)} &
\multicolumn{1}{c}{} \\
\\
\hline
\\
0009 & IC3019  & dE1,N        & $-$17.11  & 1804 & 14.04  & 30.81      & 23.48  
&
12.66  & 34.48       & 22.35  & 1.38  & a \\
0490 & IC0783  & dS0(3),N     & $-$17.18  & 1293 & 13.97  & 27.31      & 23.15  
&
12.63  & 28.35       & 21.89  & 1.34  & b \\
0781 & IC3303  & dS0(5),N:    &         & $-$254 &        &            &        
&
13.56  & 10.30       & 20.62  &       &   \\
0810 &         & dE0,N        & $-$14.47  & $-$340 & 16.68  & $\,$ 9.92  & 23.66 
 &
15.45  & $\,$ 9.94   & 22.44  & 1.23  &   \\
0815 &         & dE2, N       & $-$15.20: & $-$700 & 15.95: & 13.54:     & 
23.61: &
14.90: & 12.93:      & 22.46: & 1.05: &   \\
0846 &         & dE1,N:       & $-$14.96  & $-$730 & 16.19  & 12.35      & 23.65 
 &
14.73  & 15.35       & 22.66  & 1.46  &   \\
0850 &         & dE,N         & $-$12.63: &      & 18.52: & $\,$ 8.81: & 25.24: 
&
17.46: & $\,$ 8.27:  & 24.05: & 1.06: & c \\
0856 & IC3328  & dE1,N        & $-$16.96  &  972 & 14.19  & 20.66      & 22.76  
&
12.98  & 18.07       & 21.26  & 1.21  & d \\
0928 &         & dE6,N        & $-$15.02  & $-$254 & 16.13  & $\,$ 9.11  & 22.93 
 &
14.98  & $\,$ 8.60   & 21.65  & 1.15  &   \\
0929 & NGC4415 & d:E1,N       & $-$17.46  &  910 & 13.69  & 19.26      & 22.11  
&
12.18  & 20.72       & 20.76  & 1.51  &   \\
0940 & IC3349  & dE1,N        & $-$16.34  & 1563 & 14.81  & 19.42      & 23.25  
&
13.56  & 18.35       & 21.88  & 1.25  & b \\
0962 &         & dE3          & $-$14.18: &      & 16.97: & 22.00:     & 25.68: 
&
15.93: & 17.90:      & 24.19: & 1.04: &   \\
0998 &         & dE4,N:       & $-$13.21  &      & 17.94  & 11.37      & 25.22  
&
16.75  & 11.11       & 23.98  & 1.19  &   \\
1010 & NGC4431 & dS0(5),N     & $-$17.29  &  913 & 13.86  & 17.23      & 22.04  
&
12.47  & 17.05       & 20.63  & 1.39  & b \\
1036 & NGC4436 & dE6/dS0(6),N & $-$17.26  & 1163 & 13.89  & 16.37      & 21.96  
&
12.65  & 14.90       & 20.51  & 1.24  &   \\
1087 & IC3381  & dE3,N        & $-$17.00  &  645 & 14.15  & 19.35      & 22.58  
&
12.87  & 18.06       & 21.15  & 1.28  &   \\
1093 &         & dE0,N        & $-$14.22  &      & 16.93  & 12.23      & 24.37  
&
15.45  & 14.34       & 23.23  & 1.48  &   \\
1104 & IC3388  & dE5,N        & $-$15.66  & 1704 & 15.49  & 11.76      & 22.84  
&
14.29  & 11.30       & 21.56  & 1.20  & e \\
1129 &         & dE3          & $-$13.26  &      & 17.89  & $\,$ 7.56  & 24.28  
&
16.70  & $\,$ 7.43   & 23.05  & 1.19  &   \\
1254 &         & dE0,N        & $-$15.76  & 1350 & 15.39  & 15.11      & 23.29  
&
13.96  & 14.74       & 21.80  & 1.43  & f \\
1261 & NGC4482 & d:E5,N       & $-$17.56  & 1850 & 13.59  & 19.18      & 22.00  
&
12.38  & 18.91       & 20.76  & 1.21  &   \\
1355 & IC3442  & dE2,N        & $-$16.63  & 1332 & 14.52  & 30.80      & 23.96  
&
13.05  & 36.83       & 22.88  & 1.47  &   \\
1422 & IC3468  & E1,N:        & $-$17.35  & 1372 & 13.80  & 20.34      & 22.34  
&
12.64  & 18.95       & 21.03  & 1.16  & b \\
1567 & IC3518  & dE5/dS0(5),N &         & 1440 &        &            &        &
13.25  & 26.12       & 22.33  &       & g \\
1895 & UGC7854 & d:E6         &         & 1032 &        &            &        &
13.80  & 11.51       & 21.11  &       &   \\
\\
\hline
\end{tabular}
\end{center}
\begin{scriptsize}
Values followed by a colon have to be considered as uncertain due to disturbing
background or foreground objects. \\
$^a$ Whether this galaxy is really nucleated is not clear. The object
considered as nucleus is very faint and largely off-centered. It could 
therefore also
be a bright globular cluster which in projection happens to lie close to the 
center.
In the VCC the galaxy is classified as dE,N, whereas Miller et al. (1998) regard 
it
as non-nucleated. Stiavelli et al. (2001) present a HST observation of this 
galaxy and
provide the results of different profile fits as well as the determination of 
the
nuclear cusp slope. \\
$^b$ disk features discovered (Jerjen et al. 2001, Barazza et al. 2002) \\
$^c$ The VCC type of this galaxy, which heavily overlaps with VCC0846, is ImIV?
We suggest a reclassification as dE,N. In support of this, the galaxy 
has not been detected in the HI survey of Hoffman et al.~(1989). \\
$^d$ spiral structure discovered (Jerjen et al. 2000) \\
$^e$ As part of the Hubble Space Telescope (HST) key project the luminosity of
the tip of the red-giant branch (TRGB) for this galaxy has been determined 
(Harris et
al. 1998): $I_{TRGB} = 26.82 \pm 0.06$. The corresponding distance is $d = 15.7 
\pm
1.5$ Mpc (using $M_I^{TRGB} = -4.2 \pm 01$). \\
$^f$ A velocity profile has recently been derived by Geha et al. (2001),
showing that this galaxy is not rotating. \\
$^g$ This galaxy could not be fully integrated, since its shape is strongly
non-elliptical. The values given are estimates, using $B_T=14.52$ from Binggeli
\& Cameron (1993) and our $\langle B-R \rangle = 1.27$.
\end{scriptsize}
\end{table*}

The images were obtained using the first two units of the Very Large
Telescope (VLT) at ESO Paranal Observatory in service mode over a period of two
semesters: at UT1+FORS1 (Antu) during an observing run on July 10-14, 1999 and
at UT2+FORS2 (Kueyen) during dark time periods in March-May 2000. The detectors
of the FORS (FOcal Reducer/Low dispersion Spectrograph) instruments are thinned
and anti-reflection-coated Tektronix (FORS1) and SiTE (FORS2) CCDs with
$2048 \times 2048$ pixels. By default, service observations were taken in
standard resolution mode, with a high gain and a pixel scale of $0 \farcs 2$
pixel$^{-1}$ that yields a field of view of $6 \farcm 8 \times 6 \farcm 8$.
The CCDs were read out in the four-port mode, i.e. four amplifiers read out
one quarter of the CCD each. Three exposures of $400-600$ sec durations with
slightly different pointings were secured in each filter for each galaxy.
More details of the observations are to be reported elsewhere (Jerjen et
al., in preparation). The basic properties of the sample galaxies are listed in
Table 1. The columns are as follows:

{\em columns} (1) and (2): identifications of the galaxies; for the coordinates 
see
VCC.

{\em column} (3): morphological type in the classification system of
Sandage \& Binggeli (1984), taken from the VCC;

{\em column} (4): absolute $B$-band magnitude, based on the apparent magnitude 
given in column 6 and a mean Virgo cluster distance of 17 Mpc;

{\em column} (5): heliocentric radial velocity in km\,s$^{-1}$
(from the VCC and Binggeli et al.~1993, 
except for VCC0928, where the value of Conselice
et al.~2001 is given).

The following entries are from the photometry presented below (Sect.~4). 
We give their
meaning here as well. It should be noted that {\em all values in magnitudes are
corrected for galactic extinction}\/ using the maps of Schlegel et al. (1998).
Values with a colon are uncertain.

{\em columns} (6) and (9): total apparent magnitude in $B$ and $R$,
respectively;

{\em columns} (7) and (10): effective radius in arcsec $[\arcsec]$ in $B$ and
$R$, respectively;

{\em columns} (8) and (11): effective surface brightnesses in $B$
$[{\rm mag}/\sq\arcsec]$ and $R$ $[{\rm mag}/\sq\arcsec]$, respectively;

{\em columns} (12): total (mean) $B-R$ colour index.

The mean absolute magnitude in $B$ for all dwarfs is
$\langle M_{B_T} \rangle = -15.76$.
Taking only the 16 dwarfs originally selected into account (without the five 
faint
objects mentioned above) we get $\langle M_{B_T} \rangle = -16.42$, which is
indeed rather bright for dwarf galaxies. The mean colour of the sample is
$\langle B-R \rangle = 1.27$. This is quite blue for early-type dwarfs, but
there are three very blue outliers whose colour might be affected by other
objects: parts of VCC0962 ($B-R = 1.04$) are hidden by a bright foreground
star, VCC0815 ($B-R = 1.05$) probably has a bright background object near its
nucleus, and VCC0850 ($B-R = 1.06$) is partially overlapping with VCC0846.
Excluding these three dwarfs we get $\langle B-R \rangle = 1.30$.

\section{Photometric procedures}
The image reduction and data analysis were performed with ESO's MIDAS
package. Three exposures were available for each science field. Before
combining them, the three frames were bias-subtracted and
flat-fielded. Cosmic rays were removed by taking the median of the three
pre-reduced exposures. As the twilight flats provided did 
not lead to satisfactory results, we were forced to improve the 
flatness for some of the frames. 
The problem became obvious when we tried to do the usual way of
sky subtraction by fitting a tilted plane to sky regions on the frame 
outside the domain of the galaxies, and as unaffected as possible by 
foreground stars: strong intensity gradients remained in the background
of many images, above all in the $R$ (or $R_s$) frames, 
obviously being due to improper flat-fielding in
$R$ (or $R_s$). We believe the problem was caused by an undetected 
source of scattered light. Reliable dark sky (super) flats, which might 
have worked
better, were unfortunately impossible to construct from the science frames. 
To find a solution to the problem nonetheless, i.e.~to get 
rid of the strongest gradients, we decided to fit third-order polynomials 
to the backgrounds and then to subtract the residuals, as
is appropriate in the case of scattered light. High-order polynomial 
fitting can of course 
be quite dangerous in cases where the galaxy
covers a substantial part of the frame, leaving only little background for 
the fitting procedure. We therefore checked all polynomial fits 
for a possible contamination by the galaxy (correlation with the galaxy
position) and reduced the sky regions used where necessary.
Any remaining gradient in the background of a galaxy frame
was then finally removed 
by fitting, and subtracting, a second sky model, this time
a tilted plane (first-order polynomial) in all cases. 
The residual intensity of this second sky fit, averaged over the whole frame,
can be taken as a rough measure of the {\em systematic}\/ uncertainty in surface
brightness due to this sky flattening problem. 
The corresponding value $\langle \Delta \mu_{sky} \rangle$ [${\rm mag}/\sq\arcsec$] for
the sample galaxies is given in Table 2, $B$ and $R$ band values in
columns 3 and 4, respectively.
This means that at the radius where the $B$ or $R$ 
surface brightness profile of a galaxy
reaches the value listed, the uncertainty in surface brightness (not counting
the
usually much smaller {\em random}\/ errors shown in Fig.~1) becomes 100\%, or
0.75 mag.

Next the frames were cleaned of disturbing foreground stars or
background galaxies. The regions around the galaxies were automatically cleaned
with an algorithm designed for this purpose, whereas objects 
on the galaxy itself were erased by hand.

For the photometric zero-pointing we used standard 
stars from Landolt (1992). The 
fields
with the standard stars had been imaged several times during the observing 
nights.
We could therefore determine the zero point and extinction parameters for
each night separately.
\begin{table}[t]
\caption[]{Mean systematic uncertainties in surface brightness
due to sky flattening problems and maximum radii for full elliptical 
integration. A $\infty$ sign
indicates that the whole galaxy 
is on the frame.}
\vspace{0.3 cm}
\begin{center}
\begin{tabular}{llrrrr}
\hline
\\
\multicolumn{1}{l}{VCC} &
\multicolumn{1}{l}{Name} &
\multicolumn{1}{c}{$\langle \Delta \mu_{sky}^B \rangle$} &
\multicolumn{1}{c}{$\langle \Delta \mu_{sky}^R \rangle$} &
\multicolumn{1}{c}{$r_{max}^B$} &
\multicolumn{1}{c}{$r_{max}^R$} \\
\multicolumn{1}{l}{(1)} &
\multicolumn{1}{l}{(2)} &
\multicolumn{1}{c}{(3)} &
\multicolumn{1}{c}{(4)} &
\multicolumn{1}{c}{(5)} &
\multicolumn{1}{c}{(6)} \\
\\
\hline
\\
0009 & IC3019  & 28.09 & 27.65 & $\infty$  & 120 \\
0490 & IC0783  & 28.52 & 26.66 & 100 & 100 \\
0781 & IC3303  &       & 27.18 &     &  $\infty$  \\
0810 &         & 30.12 & 26.58 &  $\infty$  &  $\infty$  \\
0815 &         & 30.12 & 26.58 &  32 &  32 \\
0846 &         & 30.12 & 26.58 &  $\infty$  &  $\infty$  \\
0850 &         & 30.12 & 26.58 & $\infty$   &  $\infty$  \\
0856 & IC3328  & 28.47 & 27.45 & 120 & 120 \\
0928 &         & 27.77 & 26.62 &  24 &  24 \\
0929 & NGC4415 &  27.99 & 27.65 & $\infty$   & 120 \\
0940 & IC3349  & 27.77 & 26.62 &  80 &  80 \\
0962 &         & 27.77 & 26.62 &  64 & $\infty$   \\
0998 &         & 27.77 & 27.20 &  $\infty$  & $\infty$   \\
1010 & NGC4431 & 27.77 & 27.20 &  88 & 100 \\
1036 & NGC4436 & 27.77 & 27.20 &  40 &  40 \\
1087 & IC3381  & 28.14 & 27.93 &  60 &  60 \\
1093 &         & 27.14 & 27.93 &  16 &  16 \\
1104 & IC3388  & 29.97 & 27.76 &   $\infty$ &  $\infty$  \\
1129 &         & 29.97 & 27.76 &  $\infty$  &  $\infty$  \\
1254 &         & 26.51 & 25.70 &   0 &   0 \\
1261 & NGC4482 & 28.89 & 27.45 & 120 & 120 \\
1355 & IC3442  & 30.73 & 28.70 & 120 & 120 \\
1422 & IC3468  & 28.85 & 27.86 & 100 & 100 \\
1567 & IC3518  &       & 27.70 &     &  76 \\
1895 & UGC7854 &       & 26.10 &     & $\infty$   \\
\\
\hline
\end{tabular}
\end{center}
\end{table}
The center, as well as the ellipticity and position angle of the major axis for
each galaxy were determined at the isophotal level of $\sim 25 {\rm
mag}/\sq\arcsec$ by means of the ellipse fitting routine FIT/ELL3. These
parameters were then used to obtain a growth curve (integrated light profile)
by integrating the galaxy light in elliptical apertures of fixed center,
ellipticity and position angle of the major axis. Some bright galaxies in
our sample could not be integrated completely in this way 
because they had not been placed at
the center of the frame, but slightly off in order to have enough regions
for the background fitting. In these cases we integrated the {\em outer parts}\/ 
of the galaxy only in one quadrant (obviously the one opposite to the quadrant
containing the center)
and extrapolated the results to the whole galaxy.
In Table 2, columns 5 and 6 we give the equivalent ({\em maximum})
radii for the $B$ and $R$ frames,
respectively, up to which a full
integration was possible. A $\infty$-sign indicates that the whole galaxy 
is on the frame.
In the case of VCC1254 only one quadrant was used for the whole galaxy, 
as it is located very close to the giant M49.

\section{Surface photometry}
\subsection{Model-free photometric parameters and radial profiles}
\renewcommand{\dblfloatpagefraction}{0.5}
\begin{figure*}
\begin{center}
\epsfig{file=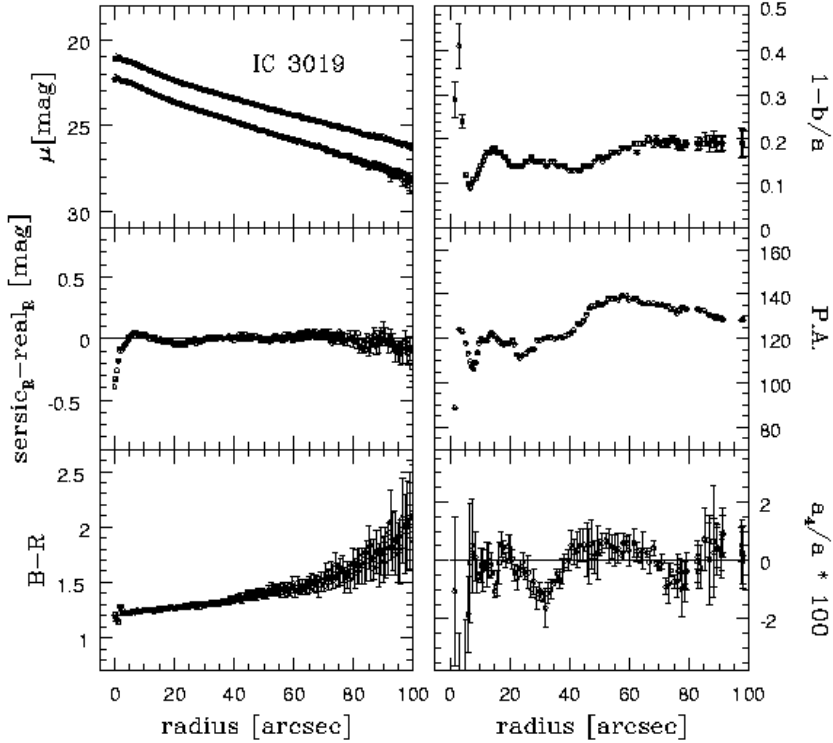,height=100mm,width=110mm}
\end{center}
\caption[]{Plots of the basic parameters for the photometry and the isophotal 
analysis
versus equivalent radius $\sqrt{ab}$, where $a$ and $b$ are the major- and 
minor-axis,
respectively. For the errors of the photometric parameters (left column) the 
remaining
gradient on the frames after the flat-fielding and the amount of the background 
subtracted
have been taken into account. The fact that in some cases only a quarter of the 
galaxy
light in the outer parts could be integrated is not considered in the error 
estimates.
Error bars are only shown for every other data point.
The errors of the isophotal parameters (right column) have
been derived by means of a Fourier expansion (for details see Bender \& 
M\"ollenhoff 1987).
Ellipses have only been fitted to isophotes completely present on the frames.
The error bars are mostly smaller than the plot symbols. \\
\vspace{0.1 cm} \\
{\bf upper left:} surface brightness profiles in $B$ (lower curve, where 
available) and
$R$ (upper curve). Error bars are only shown for the profile in $B$
(or in $R$, if $B$ is not available). A tick mark indicates the confidence limit 
of
our photometry in $R$ (Table 2, column 3), if not present, it roughly 
corresponds to
the right margin of the plot; \\
\vspace{0.1 cm} \\
{\bf middle left:} model profile (from a S\'ersic fit) minus profile observed in 
$R$; \\
\vspace{0.1 cm} \\
{\bf lower left:} $B-R$ colour-profile; \\
\vspace{0.1 cm} \\
{\bf upper right:} ellipticity profile in $R$; \\
\vspace{0.1 cm} \\
{\bf middle right:} profile of the position angle of the major-axis in $R$, from 
top
counterclockwise; \\
\vspace{0.1 cm} \\
{\bf lower right:} the profile of the isophotal shape parameter $a_4$ in $R$ 
expressed
as $a_4/a*100$, where $a$ is the length of the major-axis of the corresponding 
isophote.}
\end{figure*}
\renewcommand{\dblfloatpagefraction}{0.7}
\renewcommand{\dbltopfraction}{1}
\begin{figure*}
\begin{center}
\epsfig{file=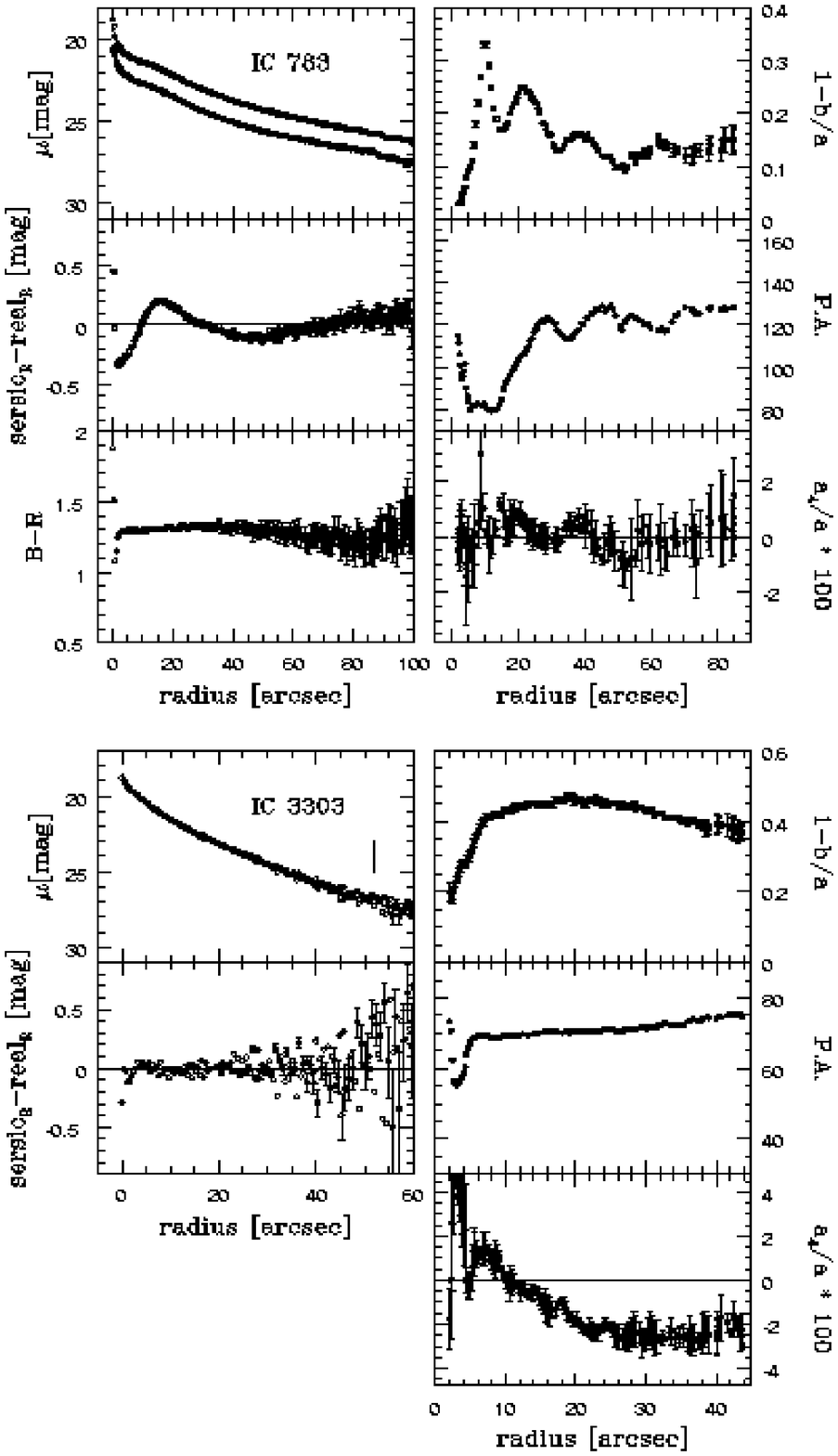,height=225mm,width=165mm}
\end{center}
{\bf Fig. 1.} continued \\
\end{figure*}
\begin{figure*}
\begin{center}
\epsfig{file=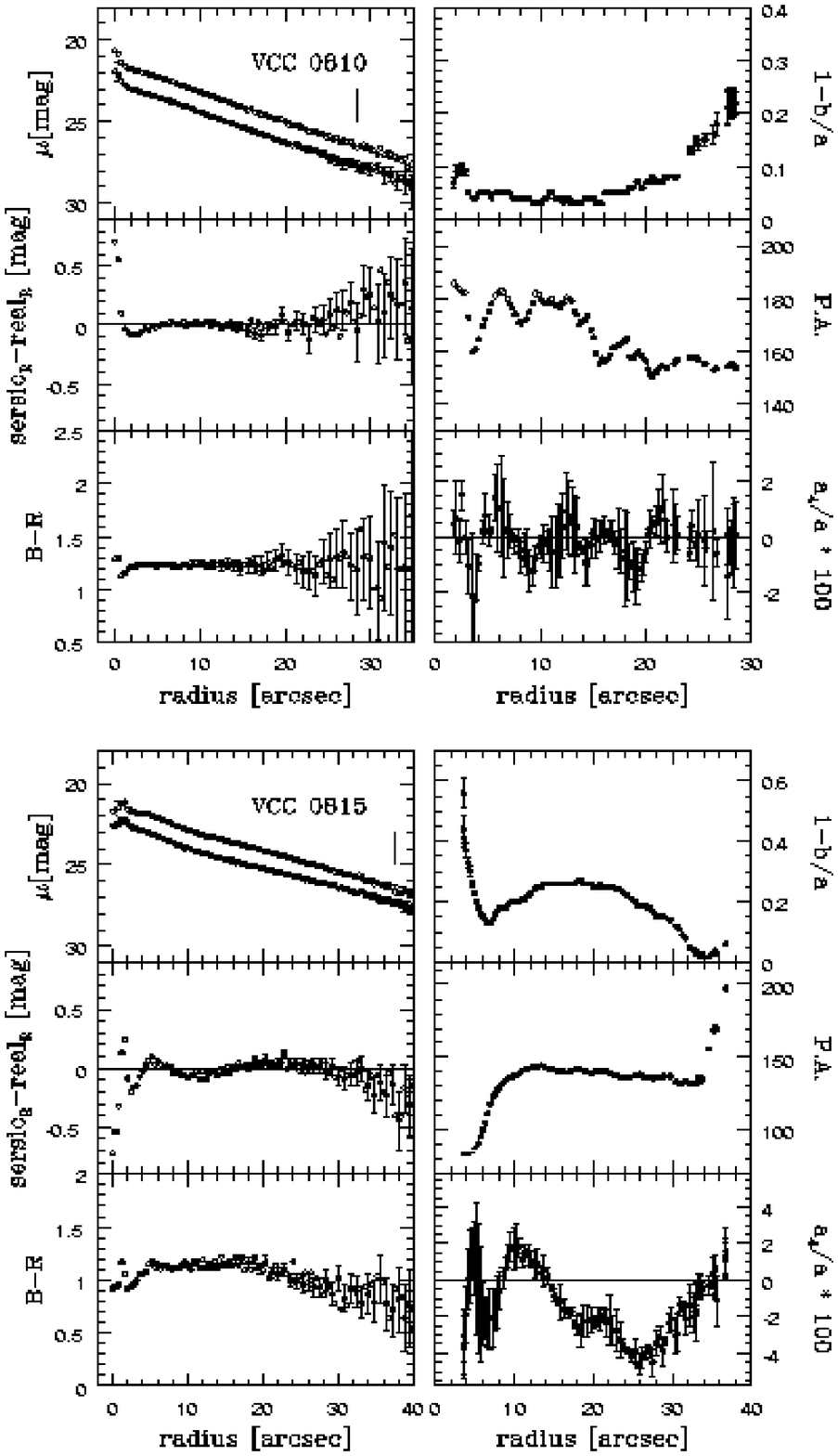,height=225mm,width=165mm}
\end{center}
{\bf Fig. 1.} continued \\
\end{figure*}
\begin{figure*}
\begin{center}
\epsfig{file=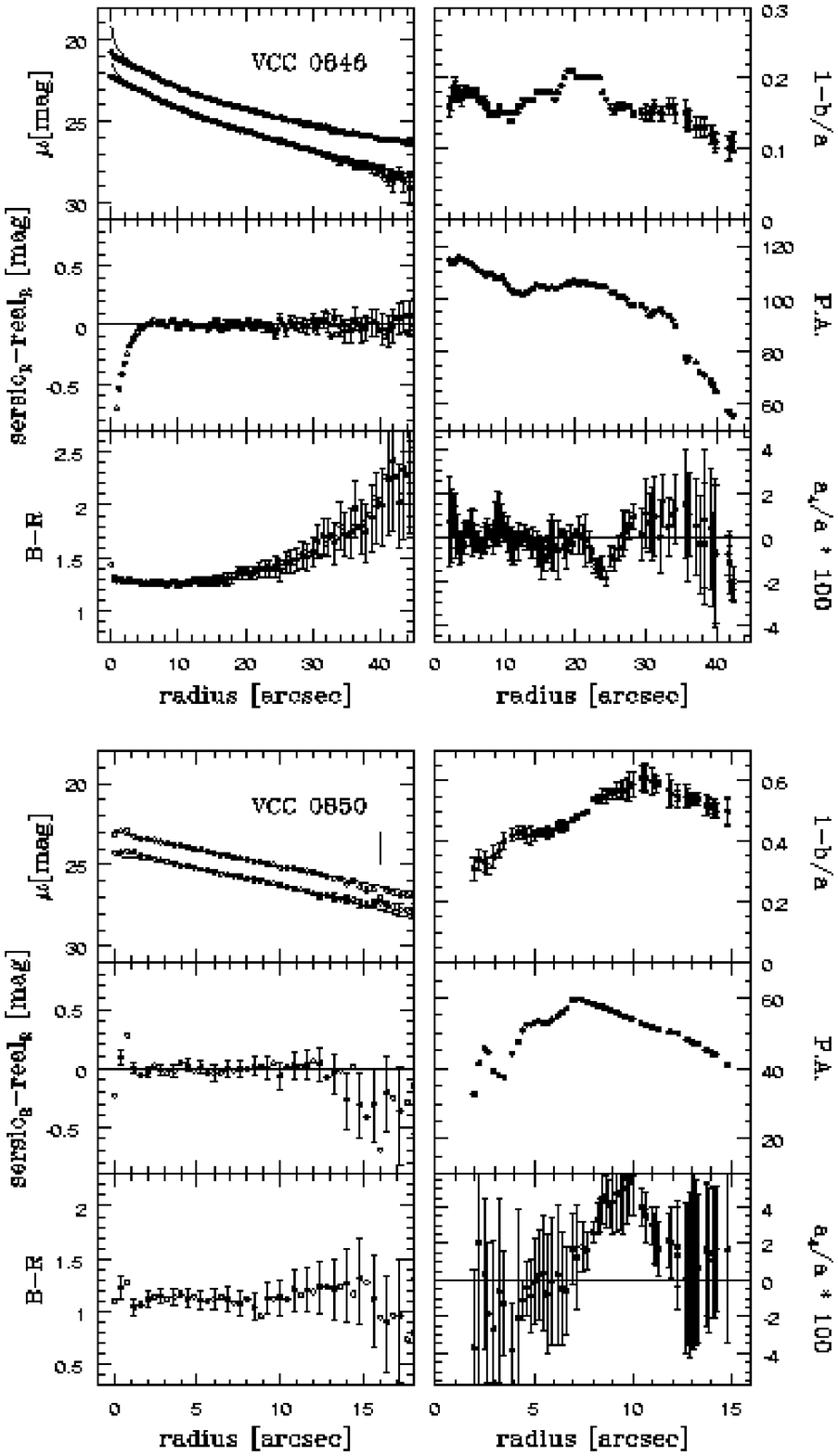,height=225mm,width=165mm}
\end{center}
{\bf Fig. 1.} continued \\
\end{figure*}
\begin{figure*}
\begin{center}
\epsfig{file=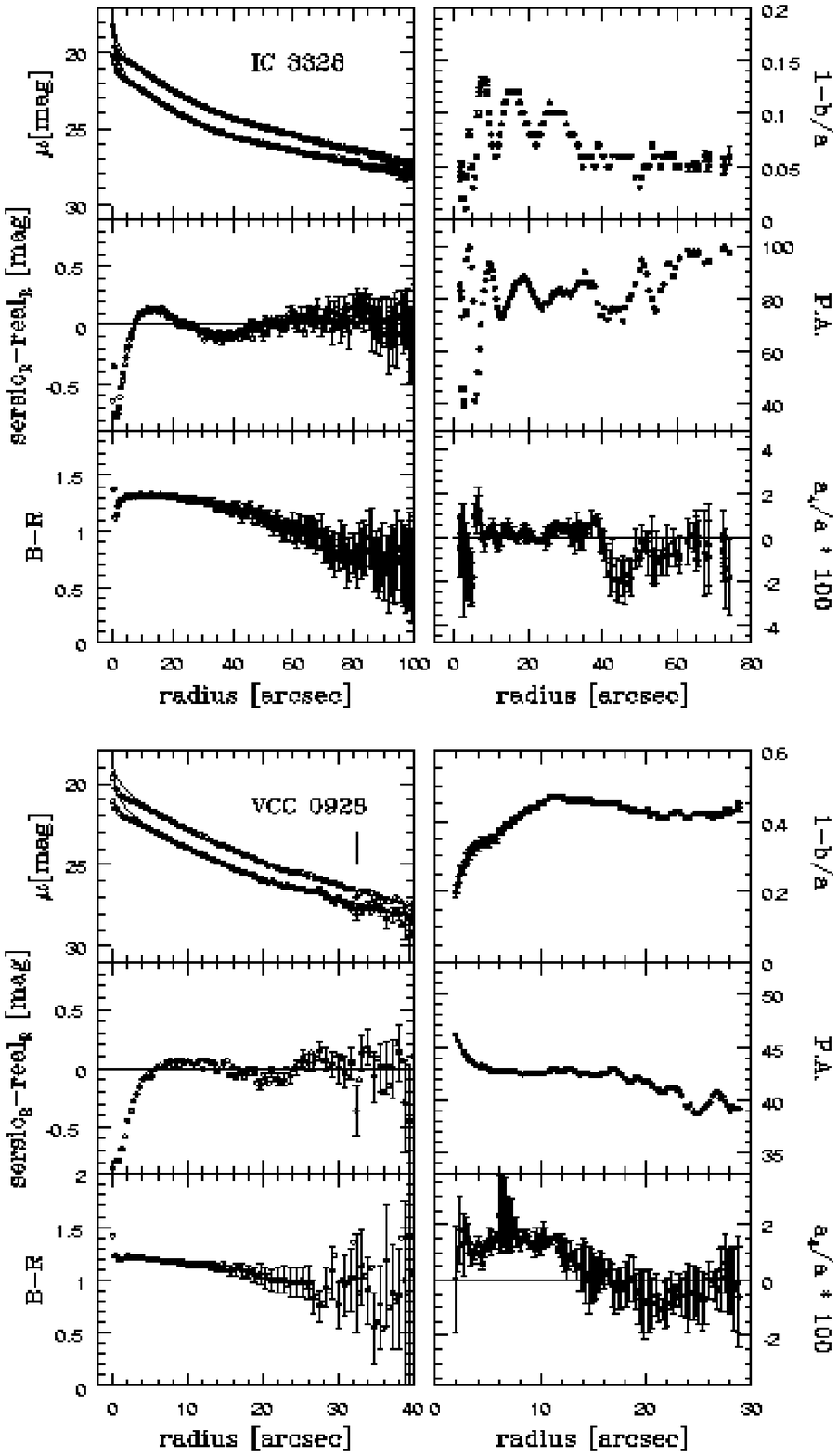,height=225mm,width=165mm}
\end{center}
{\bf Fig. 1.} continued \\
\end{figure*}
\begin{figure*}
\begin{center}
\epsfig{file=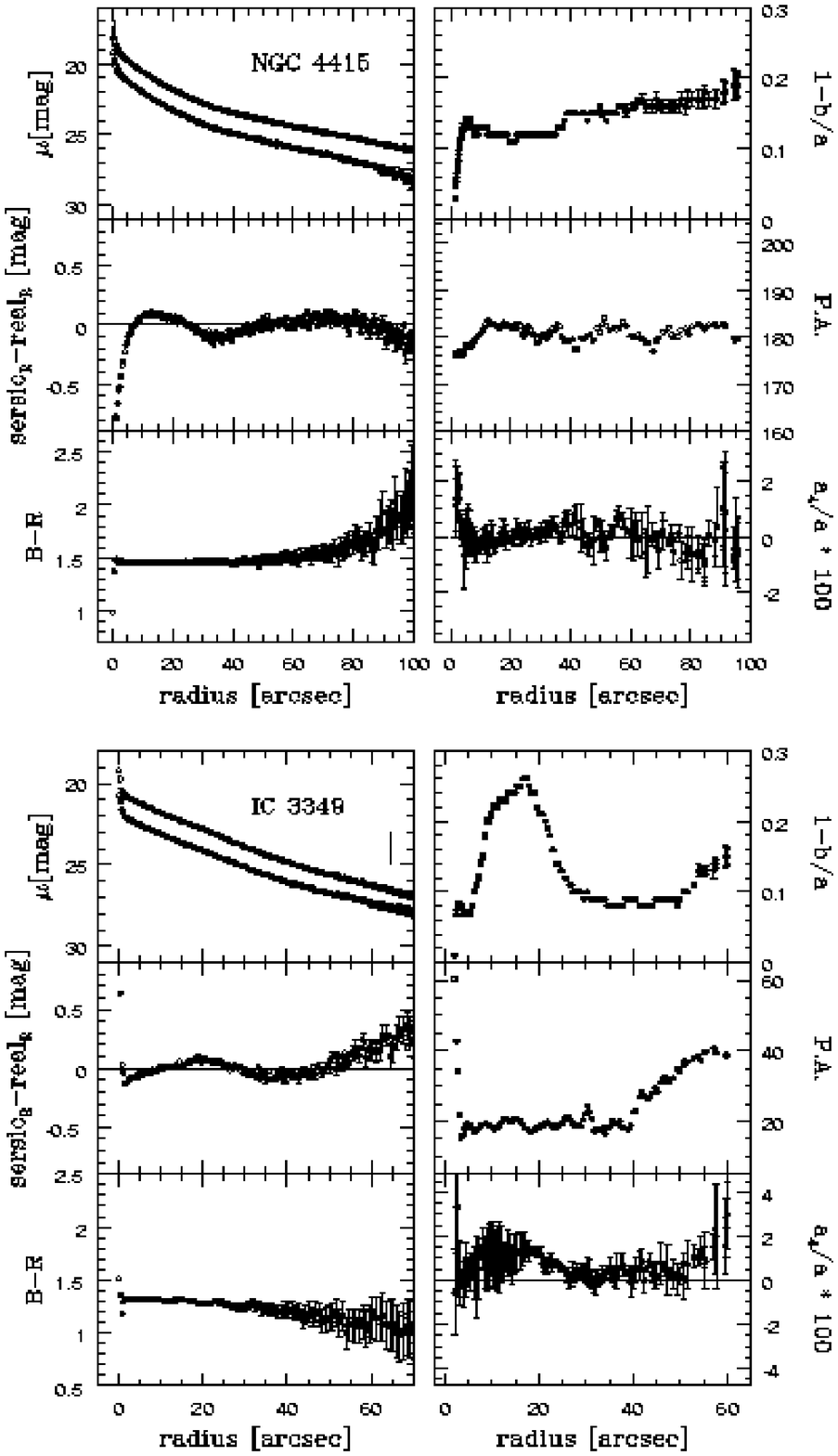,height=225mm,width=165mm}
\end{center}
{\bf Fig. 1.} continued \\
\end{figure*}
\begin{figure*}
\begin{center}
\epsfig{file=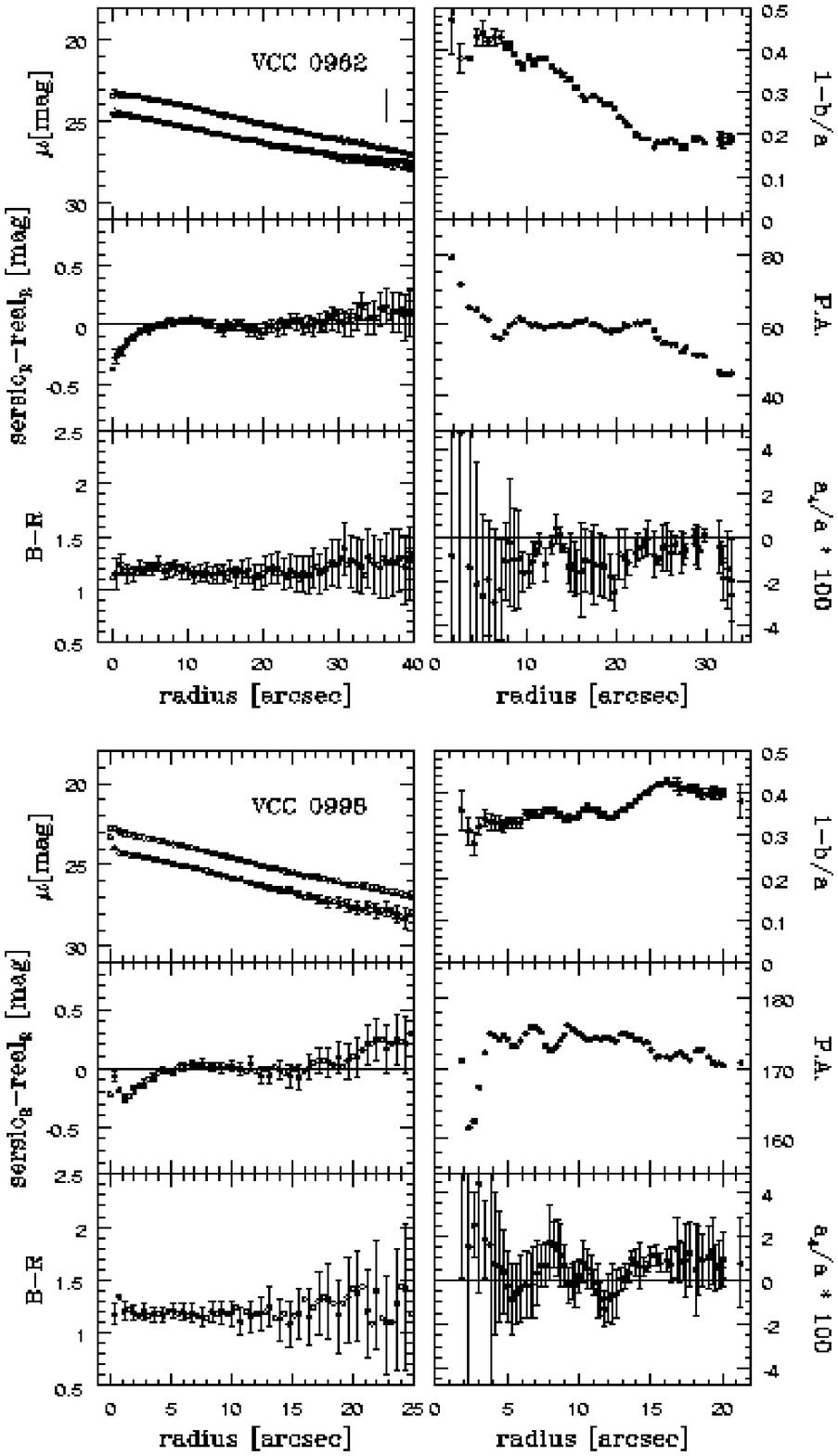,height=225mm,width=165mm}
\end{center}
{\bf Fig. 1.} continued \\
\end{figure*}
\begin{figure*}
\begin{center}
\epsfig{file=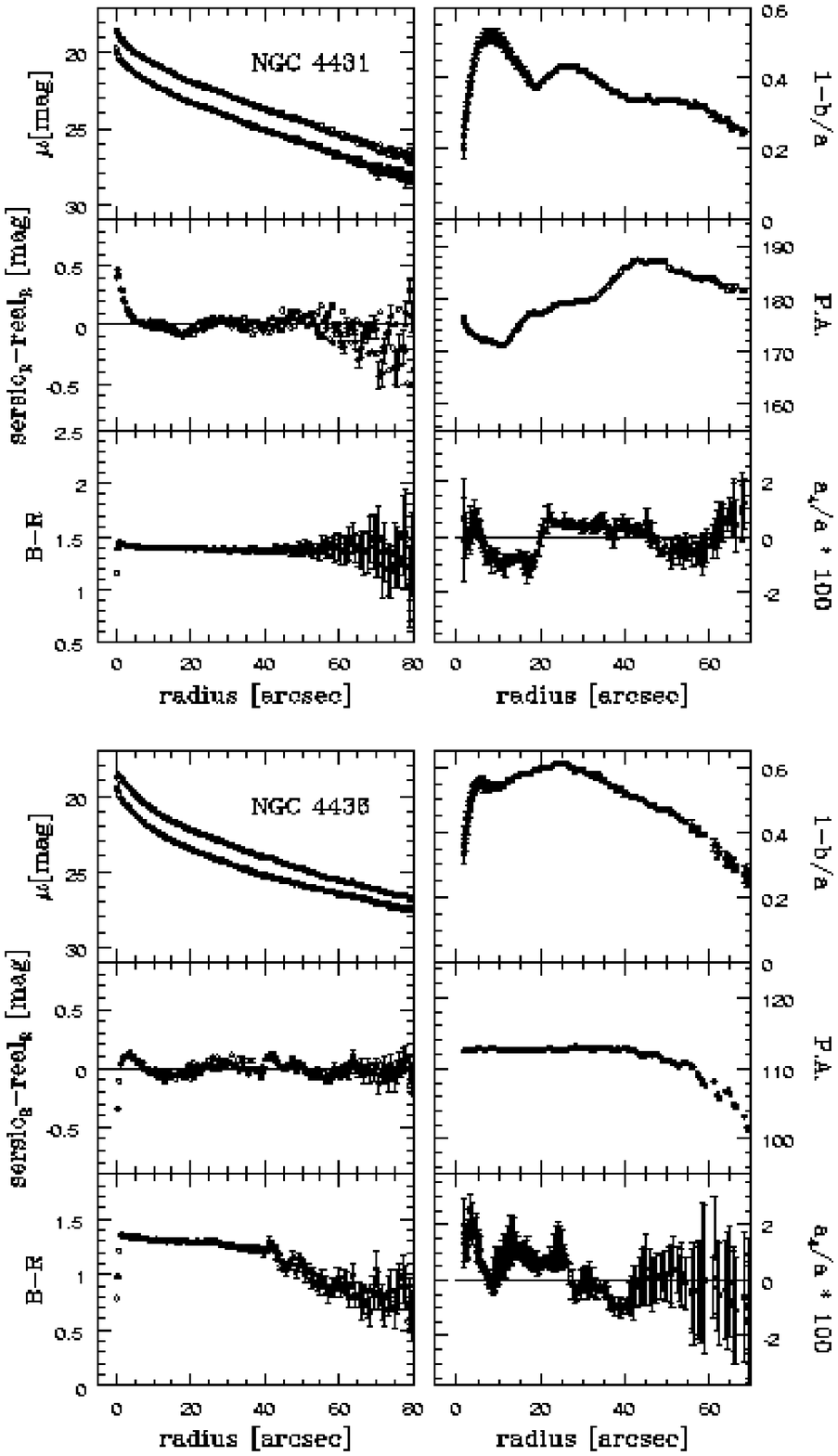,height=225mm,width=165mm}
\end{center}
{\bf Fig. 1.} continued \\
\end{figure*}
\begin{figure*}
\begin{center}
\epsfig{file=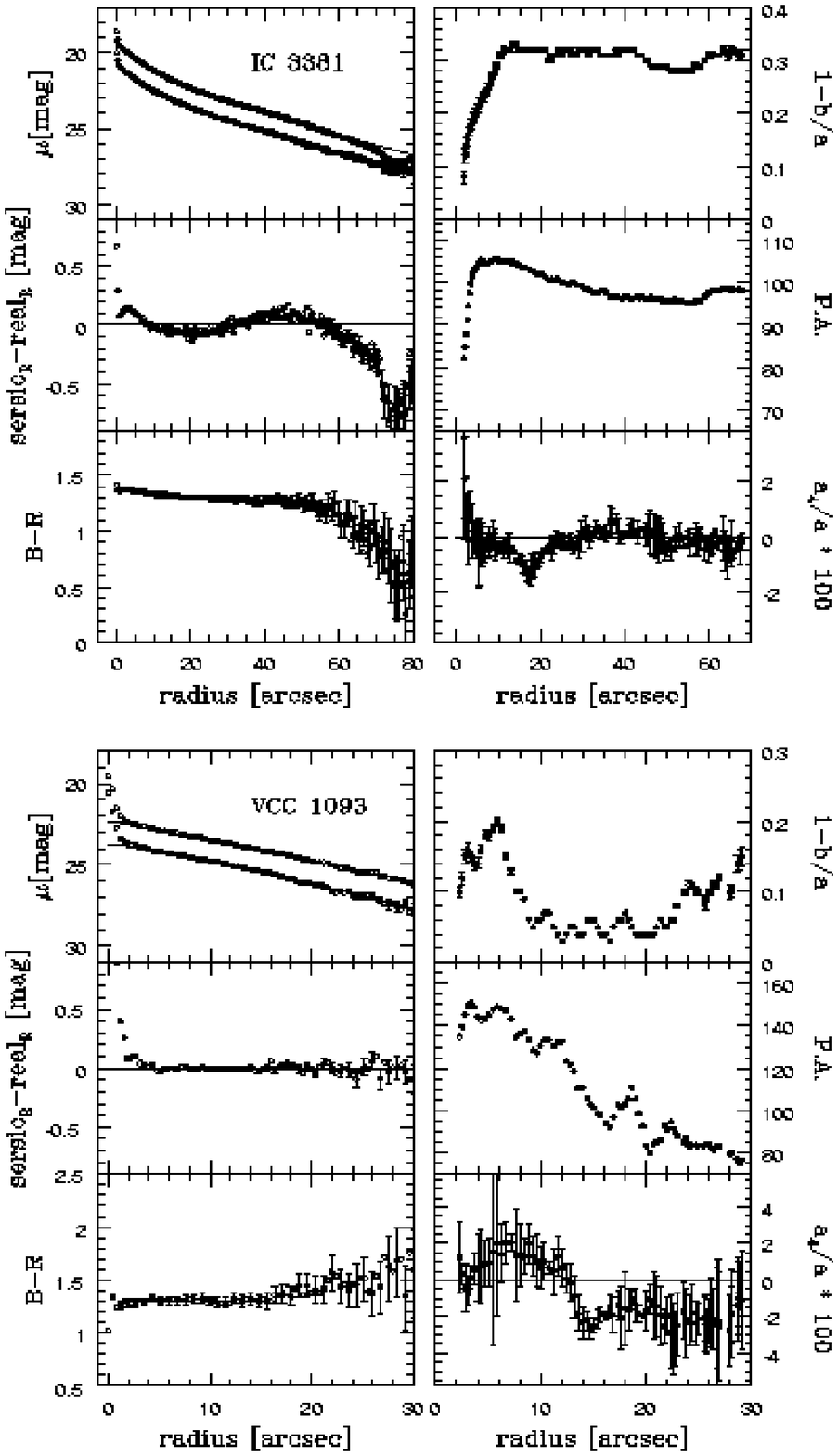,height=225mm,width=165mm}
\end{center}
{\bf Fig. 1.} continued \\
\end{figure*}
\begin{figure*}
\begin{center}
\epsfig{file=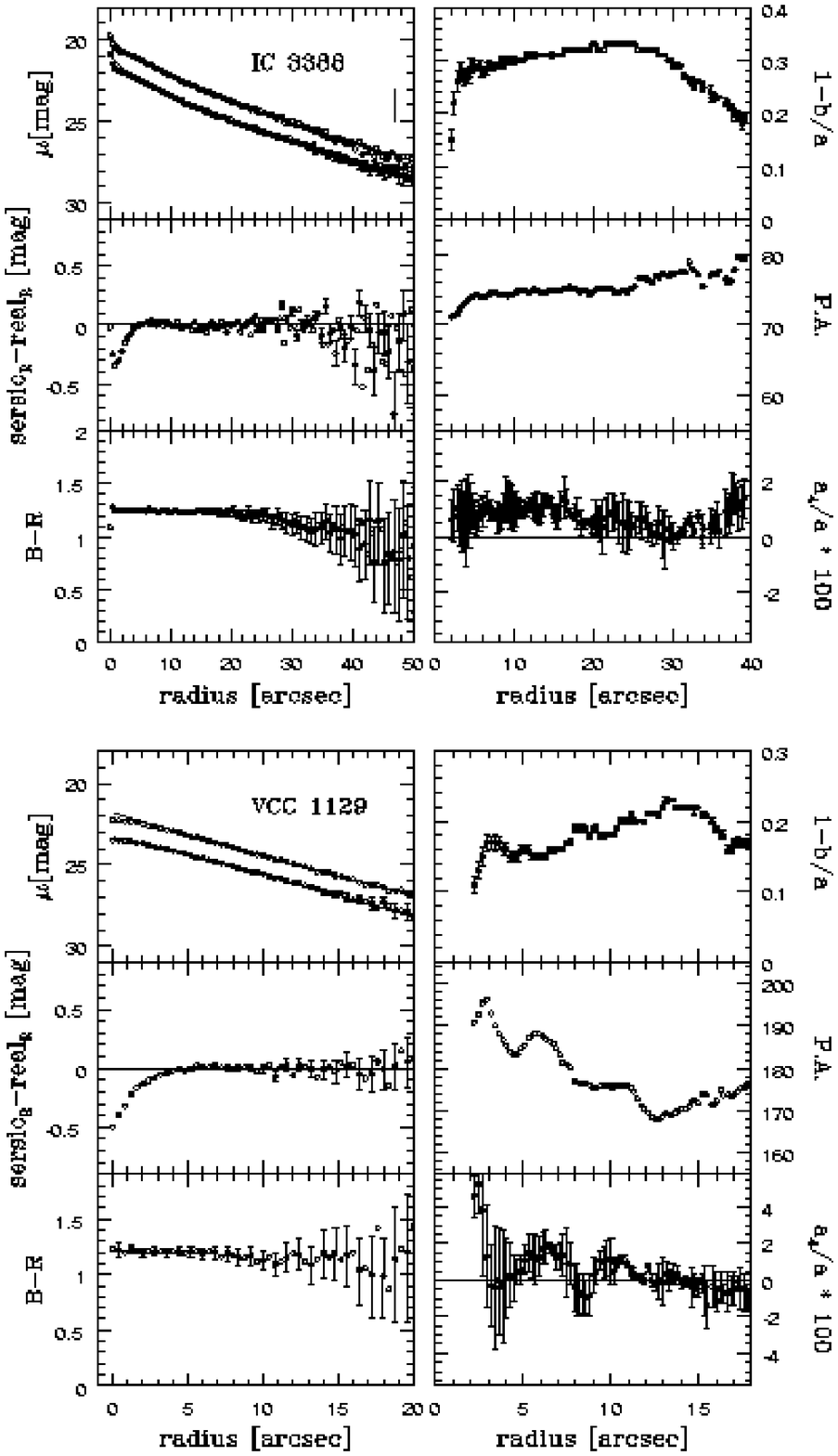,height=225mm,width=165mm}
\end{center}
{\bf Fig. 1.} continued \\
\end{figure*}
\begin{figure*}
\begin{center}
\epsfig{file=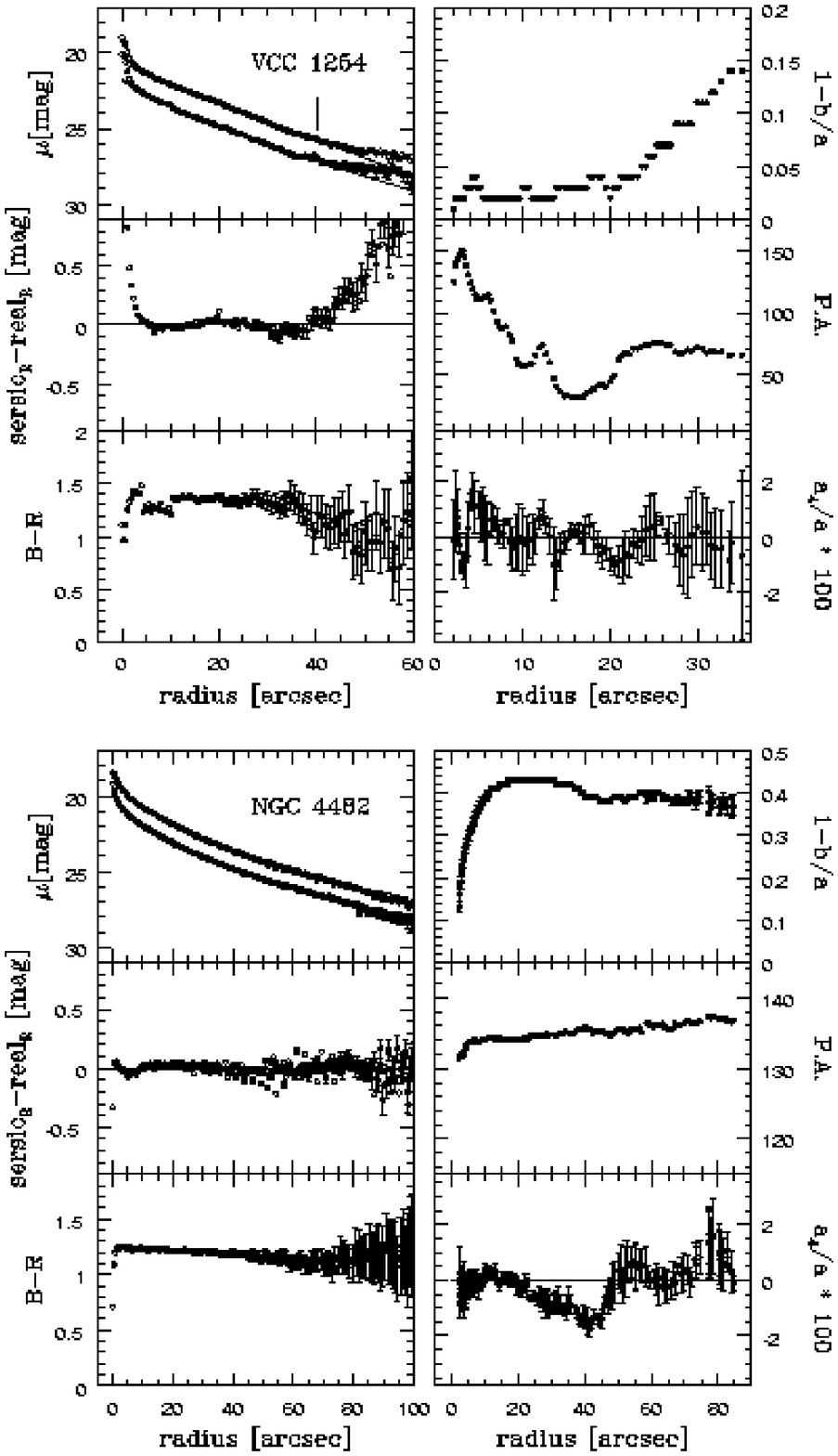,height=225mm,width=165mm}
\end{center}
{\bf Fig. 1.} continued \\
\end{figure*}
\begin{figure*}
\begin{center}
\epsfig{file=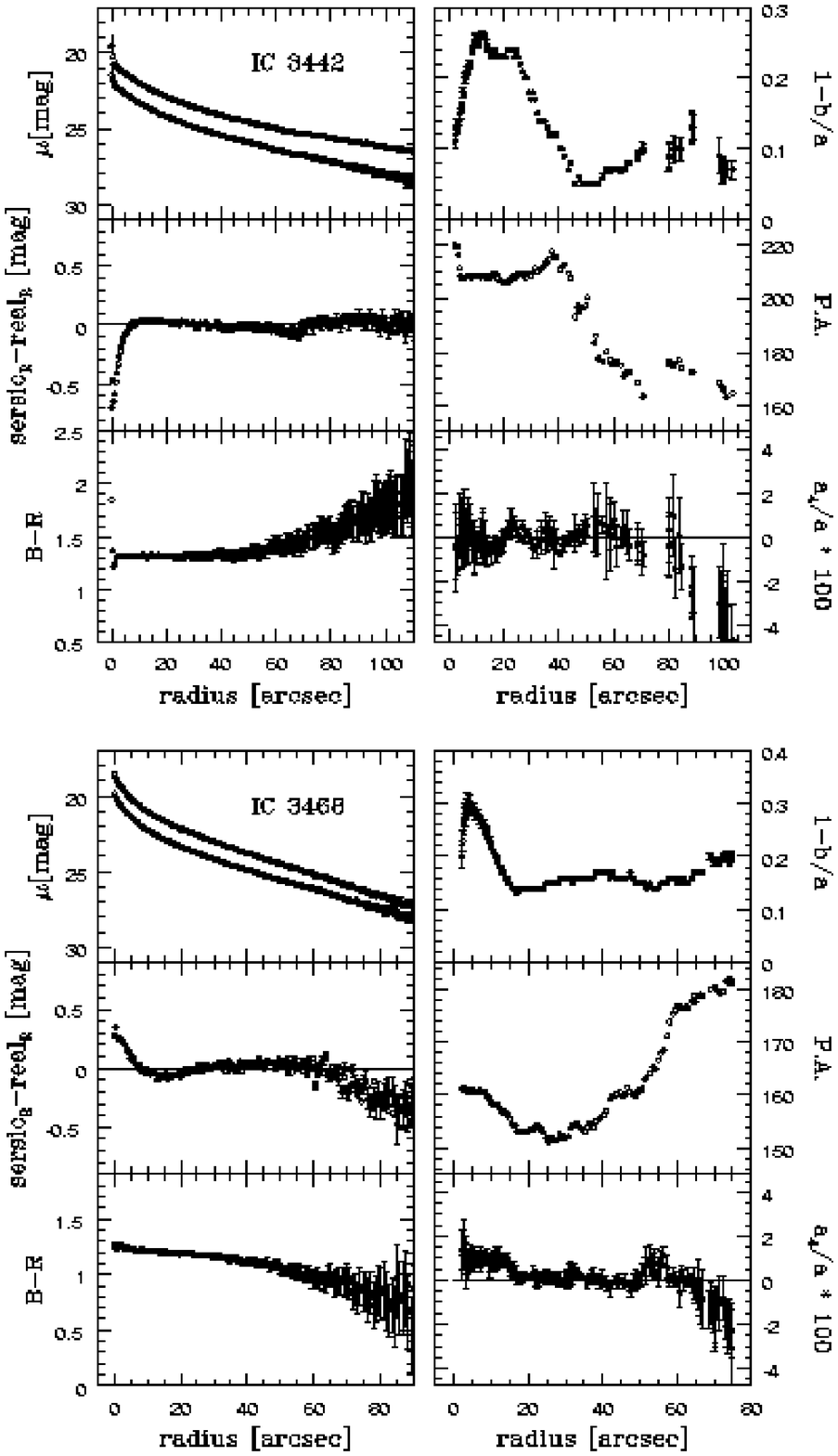,height=225mm,width=165mm}
\end{center}
{\bf Fig. 1.} continued \\
\end{figure*}
\begin{figure*}
\begin{center}
\epsfig{file=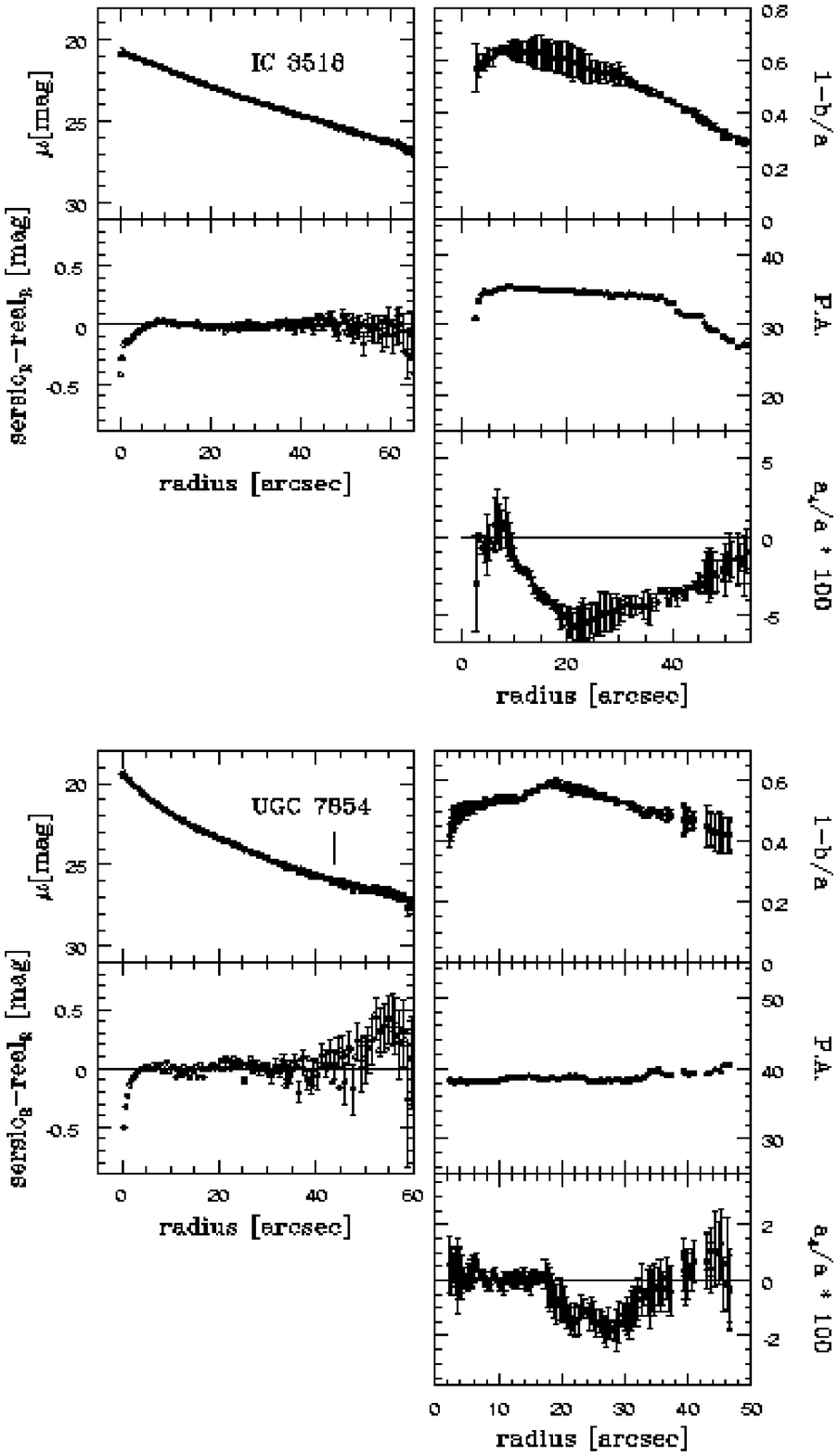,height=225mm,width=165mm}
\end{center}
{\bf Fig. 1.} continued \\
\end{figure*}
As described above, we determined the growth curve for each galaxy by
integrating the galaxy light in elliptical apertures. The intensity at which
this curve becomes asymptotically flat yields the total apparent
magnitude. In a few cases, however, the first integration promptly led to a
reasonably flat growth curve. Usually the outer shape of the growth curve shows
a continuous increase or a maximum followed by a continuous decrease, indicating
a slight, erroneous excess or
deficiency in the sky level. The shape of the curve can therefore be
used to do a fine-tuning of the sky level by simply adding or subtracting
a constant to the whole frame. Once the growth curve is corrected 
to be asymptotically flat, the effective radius can then be read off where the 
growth curve reaches
half of its asymptotic value. Using the equation
\begin{equation}
\langle\mu\rangle_{eff}\,[{\rm mag}/\sq\arcsec]=m+5\,\log(r_{eff}[\arcsec])
+1.995\,\,,
\end{equation}
where $m$ is the total apparent magnitude and $r_{eff}$ is the effective
radius, we get the effective surface brightness, $\langle \mu \rangle_{eff}$.
These model-free parameters are listed in Table 1. The meaning of the columns 
has 
already been 
given there (recall that all values are extinction-corrected).

Surface brightness profiles can be obtained by differentiating the growth curve
with respect to equivalent radius $r=\sqrt{ab}$, where $a$ and $b$ are the 
major-
and minor-axis, respectively. We used a resolution of $0\farcs4$, which
corresponds to two pixel lengths. The {\em extinction-corrected profiles}\/
are shown in Fig.~1 (left
column on top). The upper curve is the $R$ profile and the lower curve the
$B$ profile. Error bars are only shown for the $B$ profile. They can be
considered as upper limits for the profile in $R$. Only in the outer parts the
error bars are larger than the plot symbols. The errors have been estimated
using the remaining gradients on the frame after flat fielding
and the intensity of the subtracted
background. They therefore indicate the accuracy of the profile at the
corresponding surface brightness level and do not take into account the fact
that in some cases only one quarter of the galaxy light has been integrated.

In the bottom left panel of Fig.~1 we plot the $B-R$ colour profile. For dwarf
galaxies in general one would expect rather flat profiles, i.e. no strong colour
gradients. In the case of dwarf ellipticals a positive gradient, i.e. a 
reddening
towards the outer parts, could be explained by the presence of a younger
population of stars in the center, where the last star formation event took
place (Vader et al. 1988). Surprisingly, almost half of the galaxies in our
sample (10) exhibit a negative colour gradient, getting bluer towards the
outer parts. This could mean that there is a metallicity gradient in these 
galaxies.
Indeed, for five of these objects either a disk component has
been discovered or at least indications of the presence of a disk have been
found (Jerjen, Kalnajs \& Binggeli 2000; Barazza et al.~2002). 
A separate, more elaborate investigation of the colour properties
of dwarf ellipticals is in preparation (Barazza et al.~2003).

\subsection{The S\'ersic law: fits and parameters}
\renewcommand{\topfraction}{1}
\begin{figure}[t]
\begin{center}
\epsfig{file=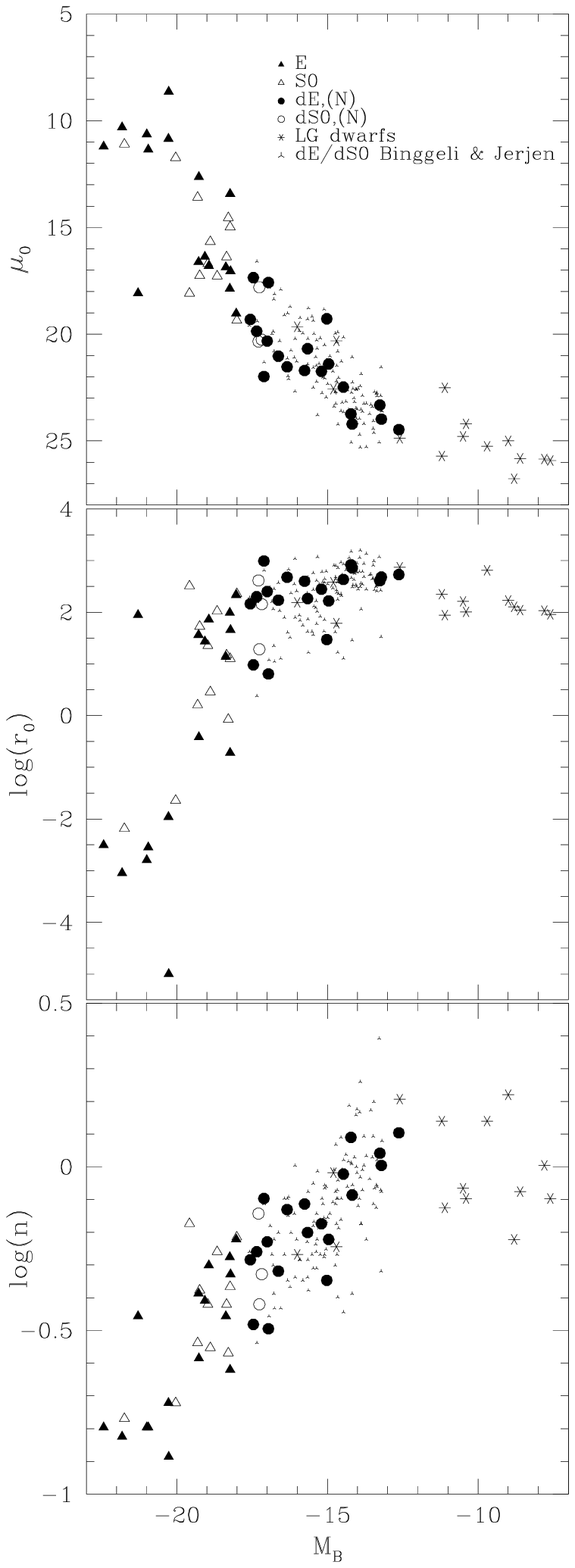,height=186mm,width=67mm}
\end{center}
\caption[]{The parameters of the best-fitting S\'ersic model, central surface
brightness ($\mu_0$), scale length ($r_0$), and shape indicator (n), plotted
versus absolute magnitude for different galaxy types. All data are in the
$B$-band. Es and S0s are from Caon et al. (1993), Local Group dSphs are from
Jerjen, Binggeli \& Freeman (2000) and dE,(N)s/dS0,(N)s are from this study. 
Added as well
is the large photographic sample of Virgo early-type dwarfs from Binggeli \&
Jerjen (1998), shown as dots.}
\end{figure}
The surface brightness profiles of dwarf galaxies, including dEs, 
can usually be fitted quite
well by an exponential model (De Vaucouleurs 1959; Binggeli \& Cameron 1991).
However, the profiles of {\em bright}\/ early-type dwarfs are known to deviate
considerably from an exponential law (Caldwell \& Bothun 1987; 
Binggeli \&
Cameron 1993). This deviation depends systematically on the luminosity of the
dwarfs. Bright objects show a shallow luminosity excess in the inner parts,
which cannot be caused by the presence of a nucleus. A better representation
of the profiles of early-type dwarfs is provided by a S\'ersic model
(S\'ersic 1968). This model, which is a simple generalization of De
Vaucouleurs' $r^{1/4}$ and exponential laws, can be written as
\begin{equation}
\sigma(r) = \sigma_0 e^{-(r/r_0)^n},
\end{equation}
where $\sigma$ is the surface brightness (intensity per area) at the equivalent
radius $r$. There are three free parameters, which are determined by a
fitting procedure: the central surface brightness $\sigma_0$, the scale length
$r_0$, and the shape parameter $n$. It is this latter parameter that takes care
of the systematic, inner deviations from an exponential described above.
The S\'ersic model has turned out to be very appropriate for dwarf ellipticals
(e.g.~Young \& Currie 1994: Binggeli \& Jerjen 1998; Ryden et al. 1999). 
We fitted our surface brightness
profiles with this model in the magnitude representation, which is
\begin{equation}
\mu(r) = \mu_0+1.086(r/r_0)^n,
\end{equation}
with $\mu_0 = -2.5\log \sigma_0+const.$ A $\chi^2_{\rm min}$ fit to the profiles 
has
been performed outside of $4\arcsec$ and above the level of
$27{\rm mag}/\sq\arcsec$ in $B$ and $26{\rm mag}/\sq\arcsec$ in $R$. This range
has been chosen in order to exclude the nuclei and to avoid the outer parts of
the galaxies which might be affected by flat-field uncertainties. The
best-fitting parameters (again {\em extinction-corrected}) are given in Table
3:

{\em column} (1) and (2): identifications of the galaxies;

{\em columns} (3) and (5): central surface brightness in $B$
$[{\rm mag}/\sq\arcsec]$ and $R$ $[{\rm mag}/\sq\arcsec]$, respectively;

{\em columns} (4) and (7): scale length (in arcsec $[\arcsec]$) in $B$ and $R$,
respectively;

{\em columns} (5) and (8): shape parameter in $B$ and $R$, respectively.

In Fig.~2 we plot these parameters versus absolute magnitude in $B$. For
comparison we have added a sample of giant ellipticals and S0s from Caon et al.
(1993), the data for the dEs and dSphs of the Local Group from Jerjen, Binggeli
\& Freeman (2000), and the large photographic sample of early-type dwarfs in 
Virgo 
(partly
coinciding with ours) from Binggeli \& Jerjen (1998). The $M_B-\mu_0$ diagram
(top panel) shows a rather tight relation comprising 
all different types of spheroidal objects. As already found and commented
upon by Jerjen \& Binggeli (1997) and Jerjen, Binggeli \& Freeman (2000), 
the early-type 
dwarfs
perfectly bridge the gap between the faint dSphs of the local group and the
giant ellipticals. The known dichotomy between Es and dEs in a
luminosity-central surface brightness plot (e.g.~Binggeli \& Cameron 1991)
vanishes in the S\'ersic representation. 
Note that the dichotomy between core and power-law
systems among normal Es (e.g.~Faber et al.~1997) also disappears here:
the S\'ersic $\mu_0$ values are not actual but {\em extrapolated}\/
central surface brightnesses from fitting the profiles {\em outside}\/ the 
central few 100 parsecs; the dichotomies mentioned are
restricted to those inner regions.

The relations in the remaining two plots are not as striking,
although in both diagrams a certain continuity between the different galaxy
classes is evident, above all among the ellipticals. In contrast, the faint
local group dSphs stand slightly apart. 

\subsection{Accuracy of the fit}
\begin{figure}[t]
\begin{center}
\epsfig{file=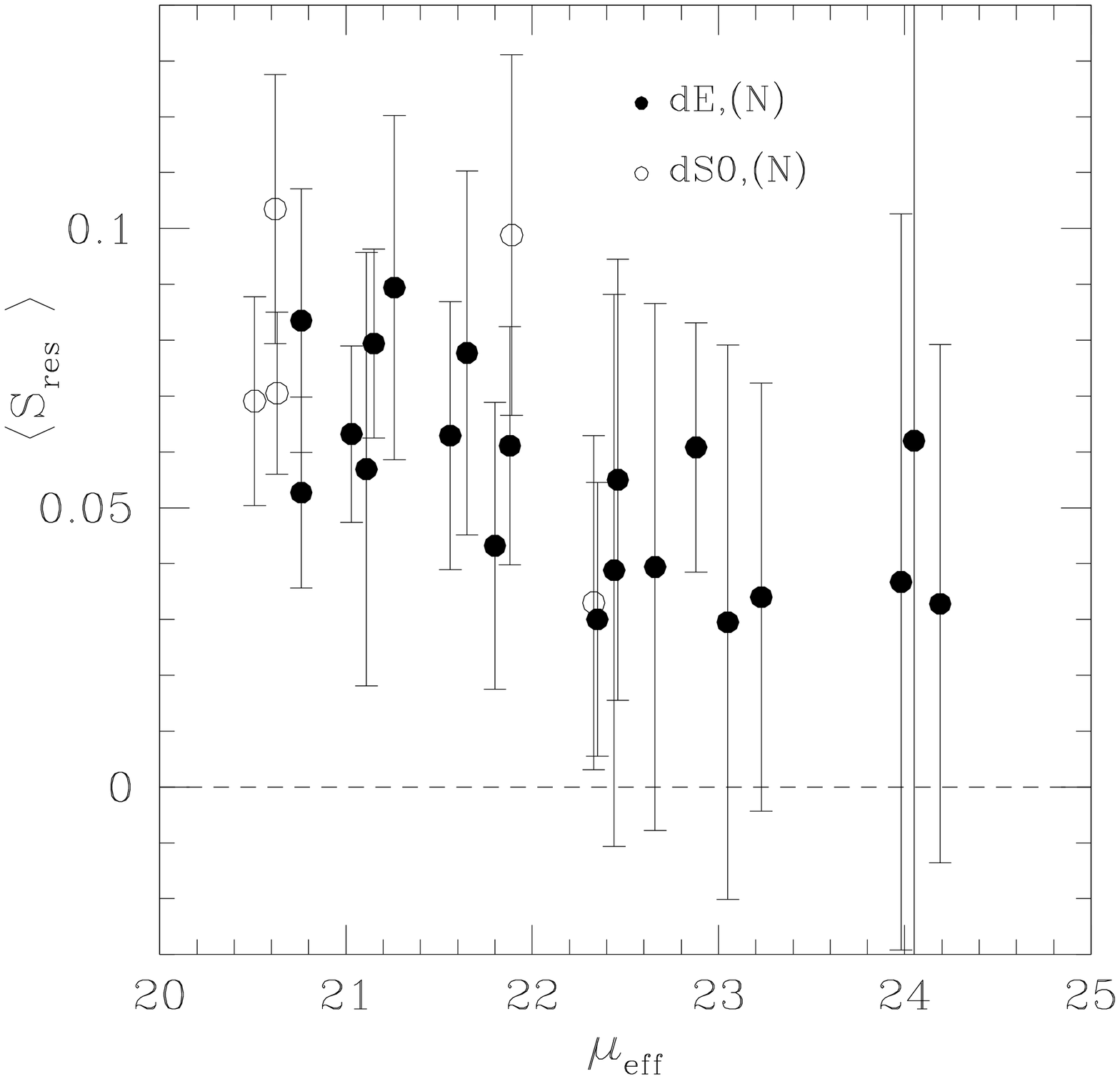,height=80mm,width=80mm}
\end{center}
\caption[]{Effective surface brightness in $R$ plotted versus the mean,
absolute residual between profile observed and model-profile (see text for
details). The error bars represent the mean error of the profile in the
corresponding range.}
\end{figure}
The best-fitting S\'ersic model profiles are plotted as solid lines through the
data points in Fig.~1 (upper left panel). The general trend of the observed
profiles is matched quite well by the models. However, in some cases the
decrease of the observed profile is not as smooth as the model in the inner
parts. This is shown in the second plot of the first column in Fig.~1, where
the difference between the observed profiles and the models is plotted. Error
bars are shown for every other data point (mostly smaller than the plot
symbols). The large residuals caused by the nuclei are not very surprising,
as the very central parts ($r <$ 4'') 
have been excluded for the fit, but a few of the bright
objects show considerable deviations from the model in the regions just outside
of the nucleus as well. A rather strong scatter is shown by the dwarfs where a
spiral structure has been discovered (IC0783, IC3328). Hence, in these cases a
bad fit might be expected. But also ``normal'' dEs, like VCC0928 or NGC4415,
show a remarkable deviation. 

In order to quantify the deviations and to find out whether they are related to
other properties of the galaxies, we determined the mean, absolute residual
between the observed profile and the fit (only for the $R$-band data):
\begin{equation}
\langle S_{res} \rangle = \sqrt{\frac{\sum_N (\mu_p-\mu_m)^2}{N}}
\end{equation}
with $N$, the number of isophote data points between $4\arcsec$ and
$26{\rm mag}/\sq\arcsec$ in $R$ (the range used for the fit), $\mu_p$, the
surface brightness measured for the isophote, $\mu_m$, the corresponding surface
brightness of the model profile. This is of course simply the standard
deviation ($1 \sigma$ scatter) of the points around the model fit.
In Fig.~3 we plot $\langle S_{res} \rangle$ versus
effective surface brightness, $\mu_{eff}$. The error bars represent the mean
error of the profile in the fitting range. Obviously, the more compact
(or brighter) objects tend to have larger residuals. Most of the faint
objects ($\mu_{eff}>22$ mag) have residuals within the errors, i.e. their
profiles are well represented by a S\'ersic model. This is not the case for
the brighter galaxies. It is important to note that the mean 
error bars shown here are much larger than the actual errors {\em in the 
central parts}\/
where the largest deviations from the model occur, i.e.~the deviations shown in
Fig.~3 are clearly significant. For dS0s the rather large
residuals might be explained by the fact that many of these galaxies are
believed to have a two-component structure, showing a high surface brightness,
lens-like feature within a more extended low surface brightness part. Hence, a
two-component model suitable for disk systems might be more appropriate for 
these
objects. However, the result remains that {\it not a single bright early-type
dwarf in our sample is well represented by the S\'ersic model}. 
Whether a second component might be present in many ``pure'' dEs as well cannot 
be answered with these data. As the deviations
mainly occur in the central parts of the profiles, a Nuker law (Lauer et al.
1995, Byun et al. 1996) or even a King model (King 1966) which are commonly
used to fit the central features of early-type
galaxies might be more suitable. However, this would increase the number of
fit-parameters and make the comparison with other objects more difficult.
Therefore, the S\'ersic model is still a useful fitting law for a general 
analysis
of the photometric properties of different object classes.
\begin{figure}[t]
\begin{center}
\epsfig{file=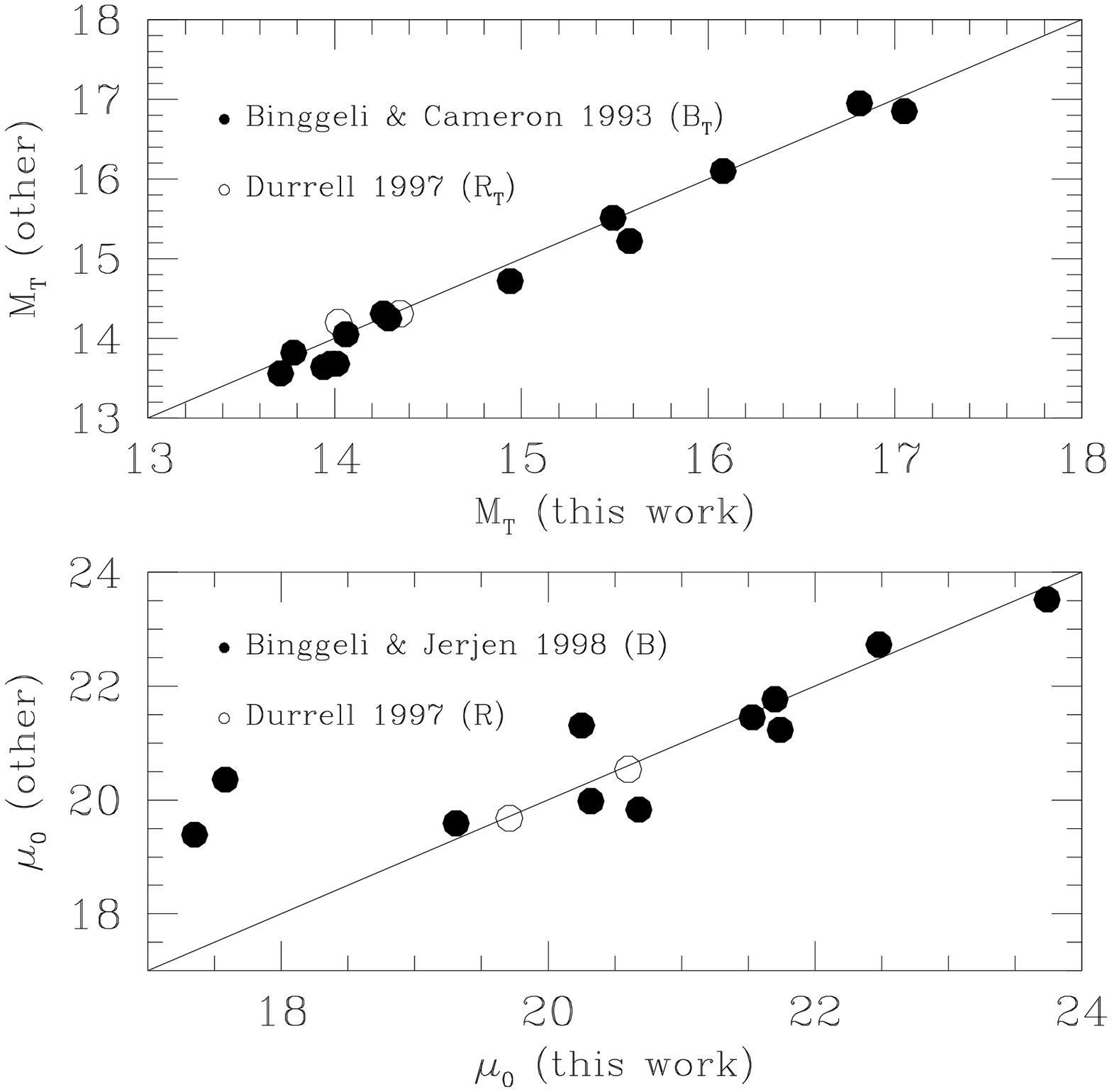,height=80mm,width=80mm}
\end{center}
\caption[]{Comparison of our total apparent magnitude in $B$ (filled circles) 
and $R$
(open circles) (upper panel) and of the central surface brightness from the 
S\'ersic fit
in $B$ (filled circles) and $R$ (open circles) (lower panel) to data from the
literature.}
\end{figure}
\begin{table*}[t]
\caption[]{Parameters of the S\'ersic fit and of the isophotal analysis.}
\vspace{0.3 cm}
\begin{center}
\begin{tabular}{llrrrrrr|rrrrr}
\hline
& & & & & & & & & & & \\
\multicolumn{1}{l}{VCC} &
\multicolumn{1}{l}{Name} &
\multicolumn{1}{c}{$\mu_0^B$} &
\multicolumn{1}{c}{$r_0^B$} &
\multicolumn{1}{c}{$n^B$} &
\multicolumn{1}{c}{$\mu_0^R$} &
\multicolumn{1}{c}{$r_0^R$} &
\multicolumn{1}{c}{$n^R$} &
\multicolumn{1}{|c}{$\langle \epsilon \rangle$} &
\multicolumn{1}{c}{$\langle P.A. \rangle$} &
\multicolumn{1}{c}{$\Delta P.A.$} &
\multicolumn{1}{c}{$\langle \frac{a_4}{a}*100 \rangle$} &
\multicolumn{1}{c}{$\langle \delta r_N \rangle$} \\
\multicolumn{1}{l}{(1)} &
\multicolumn{1}{l}{(2)} &
\multicolumn{1}{c}{(3)} &
\multicolumn{1}{c}{(4)} &
\multicolumn{1}{c}{(5)} &
\multicolumn{1}{c}{(6)} &
\multicolumn{1}{c}{(7)} &
\multicolumn{1}{c}{(8)} &
\multicolumn{1}{|c}{(9)} &
\multicolumn{1}{c}{(10)} &
\multicolumn{1}{c}{(11)} &
\multicolumn{1}{c}{(12)} &
\multicolumn{1}{c}{(13)} \\
& & & & & & & & & & & \\
\hline
& & & & & & & & & & & \\
0009 & IC3019  & 21.98 & 12.25 & 0.80 & 20.71 & 11.92 & 0.76 &
0.16 & 123.7 & 33.0 & $-$0.29$\quad$ & 2.45 \\
0490 & IC0783  & 20.25 &  1.80 & 0.47 & 19.17 &  2.49 & 0.51 &
0.18 & 106.8 & 50.2 & $\,$0.20$\quad$ & 0.78 \\
0781 & IC3303  &       &       &      & 18.52 &  2.02 & 0.63 &
0.42 &  69.4 & 15.6 & $-$0.43$\quad$ & 0.34 \\
0810 &         & 22.48 &  5.38 & 0.95 & 21.33 &  5.80 & 0.99 &
0.05 & 139.0 &  4.9 & $-$0.08$\quad$ & 0.42 \\
0815 &         & 21.74 &  3.50 & 0.67 & 20.96 &  5.18 & 0.80 &
0.21 & 131.7 & 60.5 & $-$1.08$\quad$ & 0.79 \\
0846 &         & 21.39 &  2.07 & 0.60 & 19.14 &  0.55 & 0.43 &
0.17 & 103.7 & 43.2 & $-$0.05$\quad$ & 0.63 \\
0850 &         & 24.47 &  6.69 & 1.27 & 22.92 &  4.70 & 0.93 &
0.51 &  54.2 & 12.3 & $\,$ 1.66$\quad$ & 0.45 \\
0856 & IC3328  & 17.58 &  0.08 & 0.32 & 17.59 &  0.54 & 0.42 &
0.09 &  80.7 & 29.0 & $-$0.05$\quad$ & 0.20 \\
0928 &         & 19.28 &  0.37 & 0.45 & 18.83 &  0.92 & 0.55 &
0.42 &  42.5 &  4.5 & $\,$ 0.99$\quad$ & 0.19 \\
0929 & NGC4415 & 17.35 &  0.12 & 0.33 & 15.59 &  0.07 & 0.31 &
0.13 &  60.2 &  8.1 & $-$0.01$\quad$ & 0.26 \\
0940 & IC3349  & 21.53 &  5.95 & 0.74 & 20.25 &  6.30 & 0.77 &
0.16 &  20.2 & 22.3 & $\,$ 0.79$\quad$ & 0.37 \\
0962 &         & 24.21 &  9.03 & 0.82 & 23.05 & 10.27 & 0.98 &
0.31 &  59.1 & 12.8 & $-$1.10$\quad$ & 0.78 \\
0998 &         & 23.98 &  5.99 & 1.01 & 22.56 &  4.95 & 0.89 &
0.36 & 174.0 &  4.9 & $\,$ 0.31$\quad$ & 0.56 \\
1010 & NGC4431 & 20.34 &  5.17 & 0.72 & 18.94 &  5.15 & 0.72 &
0.43 & 140.5 & 16.1 & $-$0.20$\quad$ & 0.46 \\
1036 & NGC4436 & 17.80 &  0.24 & 0.38 & 17.76 &  1.29 & 0.51 &
0.55 & 112.5 &  7.3 & $\,$ 0.52$\quad$ & 0.91 \\
1087 & IC3381  & 20.32 &  3.15 & 0.59 & 19.25 &  4.31 & 0.66 &
0.30 & 101.4 & 10.7 & $-$0.31$\quad$ & 0.21 \\
1093 &         & 23.74 & 10.40 & 1.23 & 22.33 &  9.61 & 1.09 &
0.08 & 117.1 & 67.7 & $-$0.36$\quad$ & 0.52 \\
1104 & IC3388  & 20.68 &  2.30 & 0.63 & 19.71 &  3.13 & 0.71 &
0.30 &  74.9 &  5.3 & $\,$ 0.83$\quad$ & 0.27 \\
1129 &         & 23.32 &  5.11 & 1.10 & 21.76 &  3.76 & 0.93 &
0.18 &  90.1 & 20.4 & $\,$ 0.59$\quad$ & 0.21 \\
1254 &         & 21.70 &  4.99 & 0.77 & 20.60 &  7.21 & 0.91 &
     &       &      &       &      \\
1261 & NGC4482 & 19.31 &  1.82 & 0.52 & 18.14 &  2.00 & 0.54 &
0.40 & 134.5 &  4.6 & $-$0.27$\quad$ & 0.40 \\
1355 & IC3442  & 21.03 &  2.14 & 0.48 & 18.87 &  0.58 & 0.37 &
0.16 &  56.7 & 44.5 & $-$0.10$\quad$ & 1.93 \\
1422 & IC3468  & 19.86 &  2.49 & 0.55 & 18.81 &  3.12 & 0.60 &
0.18 & 157.1 & 29.3 & $\,$ 0.41$\quad$ & 0.33 \\
1567 & IC3518  &       &       &      & 20.43 &  7.52 & 0.81 &
0.56 &  34.2 &  8.6 & $-$3.16$\quad$ & 1.37 \\
1895 & UGC7854 &       &       &      & 18.50 &  1.29 & 0.55 &
0.54 &  38.5 &  2.0 & $-$0.25$\quad$ & 0.37 \\
& & & & & & & & & & & \\
\hline
\end{tabular}
\end{center}
\end{table*}
\subsection{Comparison with data from the literature}
In the upper panel of Fig.~4 we compare our total magnitudes with available 
data
from the literature. Binggeli and Cameron (1993) give total apparent magnitudes 
in $B$
for a large sample of Virgo dEs. The 14 objects in common with our sample are 
shown as
filled circles. In general, the data agree quite well. However, there are 
significant
differences for some galaxies, which amount to $0.3$ mag or more in four cases. 
Only
two objects can be compared with the data from Durrell (1997) in the $R$-band 
(open
circles). It is evident from Fig.~4 (upper panel) that our magnitudes tend to 
be slightly
fainter than the ones from the literature. This could indicate that we have
overestimated the background in some frames, due to the presence of
several galaxies on the 
frame,
leaving only a small region of the sky. Overall, we estimate the errors of the 
total
apparent magnitudes to be $0.15$ mag in both filters.

In the lower panel of Figure 4 we compare our central surface brightnesses in 
$B$ and
$R$, derived by a S\'ersic fit, to the data of Binggeli and Jerjen (1998) and 
Durrell
(1997), respectively. Obviously, the two brightest objects strongly disagree. 
The fact
that Binggeli and Jerjen used growth-curves for the fitting procedure cannot 
account
for these differences (see their discussion of this point), nor could 
a difference in the seeing play a role here, as the S\'ersic law fitting
is done outside the central 3$\arcsec$ or 4$\arcsec$. However, since 
brighter
objects usually have a steeper rise of the profile in the central parts, only 
slightly
different S\'ersic fits can have large deviations of the central surface 
brightness. Also, the brightest objects show generally the strongest deviations
from the S\'ersic law (see 4.3 above).

\section{Isophotal analysis}
Thanks to the high resolution of our observations we were able to perform an 
accurate
analysis of the isophotal properties of the sample galaxies. To determine the
corresponding parameters, we fitted ellipses to the isophotes, using a method 
described
by Bender \& M\"ollenhoff (1987), which is implemented in MIDAS as FIT/ELL3. In
addition, we derived the deviations of the isophotes from pure ellipses and 
expanded
them in a Fourier series. The fourth cosine parameter, $a_4$, indicates the 
shape of
the corresponding isophote: an isophote with a negative $a_4$ is called 
``boxy'',
otherwise it is called ``disky'' (see Bender \& M\"ollenhoff (1987) for 
details). In
this way, we determined the ellipticities, the position angles of the major-axis
and the shape parameters for all isophotes between $r \approx 2\arcsec$ and $\mu
\approx 26.5$ mag in $R$ with an interval of $0.05$ mag for each galaxy. In 
Fig.~1,
right column, we plot these three parameters versus equivalent radius. The error 
bars
have been determined using the prescription of Bender \& M\"ollenhoff (1987). 
The
ellipticity is defined as $\epsilon = 1-b/a$ where $a$ and $b$ are the major- 
and
minor-axis respectively, and $a_4$ is given as $a_4/a*100$.
\renewcommand{\dblfloatpagefraction}{0.5}
\begin{figure*}
\begin{center}
\epsfig{file=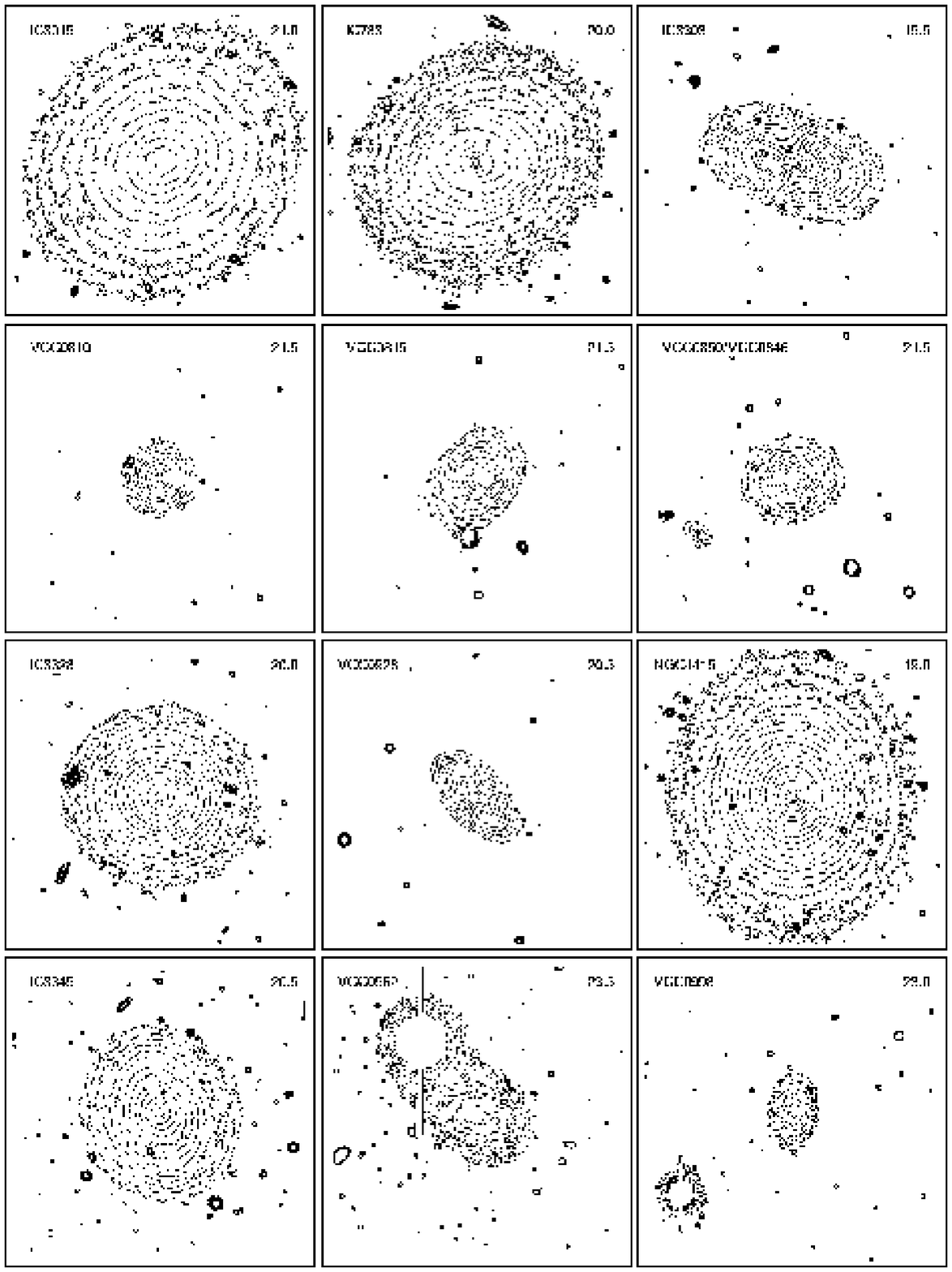,height=220mm,width=165mm}
\end{center}
\caption[]{Isophotal contours of the $R$-band images. The galaxy name is given 
in the
upper left corner. The surface brightness of the innermost isophote is indicated 
in
the upper right corner, the interval is 0.25 mag. The images are $1 \farcm 7$ on 
a
side, corresponding to $\sim 8$ kpc.}
\end{figure*}
\renewcommand{\dblfloatpagefraction}{0.7}
\renewcommand{\dbltopfraction}{1}
\begin{figure*}
\begin{center}
\epsfig{file=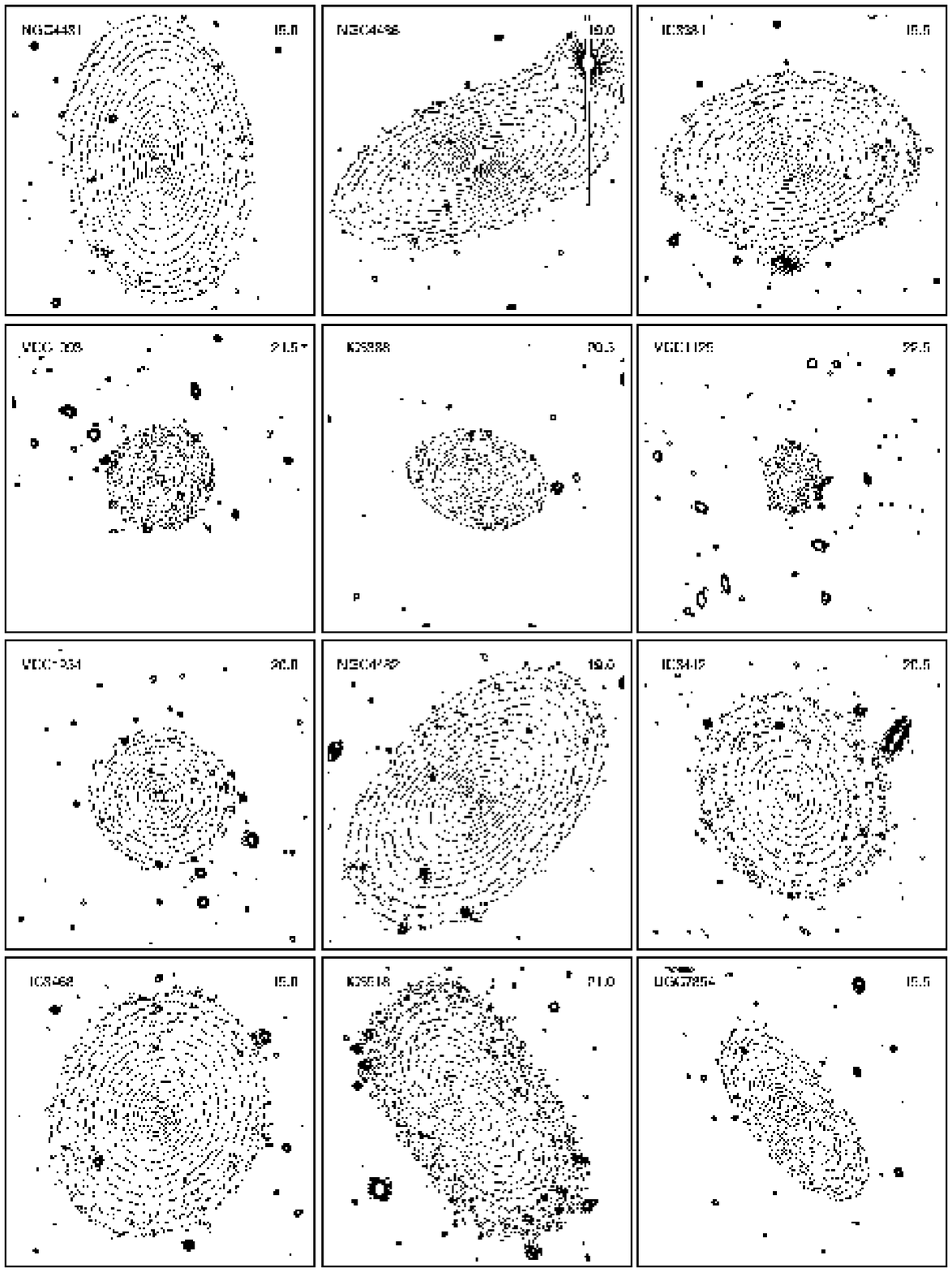,height=220mm,width=165mm}
\end{center}
{\bf Fig. 5.} continued \\
\end{figure*}

Since the center of the ellipse searched for is a free parameter in the fitting 
procedure,
we could simultaneously measure the distance, $\delta_N$, between the nucleus
(or the center of the innermost isophote for not-nucleated dwarfs) and the 
center
of the overall brightness distribution. This allows to determine a possible
off-center position of the nucleus or the concentricity of the successive 
isophotes.
The isophotal contours of the sample galaxies are shown in Fig.~5.

In order to investigate whether there are correlations between the isophotal
parameters and other properties of the galaxies, we determined their mean, 
global values.
The range within which the averaging was performed was again the same as for the
profile fitting: outside the central $4 \arcsec$ and above an isophotal level
of $26{\rm mag}/\sq\arcsec$. Following the procedures of Ryden et al.~(1999),
the parameters of the isophotes contributing to the
mean have been weighted by the fraction of intensity corresponding to that 
isophote. In this way the faint outer regions, where the errors are large,
are automatically given less weight.
Hence, the intensity - or luminosity - weighted mean is:
\begin{equation}
\langle z \rangle = \frac{\sum z dI}{\sum dI}
\end{equation}
where $z$ represents $\epsilon$, $P.A.$, $a_4$ or $\delta r_N$, and $dI$ is the
contribution of intensity of the isophote. In addition to these mean values, we
determined the maximal isophotal twist of the galaxies within the considered
radial range. The parameters are listed in Table 3, where the columns are:

{\em column} (9): ellipticity defined as $\epsilon = 1-b/a$;

{\em column} (10): position angle of the major-axis, measured from top
counterclockwise;

{\em column} (11): maximal isophotal twist;

{\em column} (12): isophotal shape parameter given as $\frac{a_4}{a}*100$;

{\em column} (13): nuclear off-set in arcsec, if no nucleus is present, the
innermost isophote is taken as reference.

We omit VCC1254 in this analysis, since its isophotal parameters can not be
determined with the required accuracy, due to its proximity to M49.

\subsection{The isophotal shape parameter $a_4$}
Giant ellipticals as well as dwarf ellipticals owe their name to the fact that 
the
shape of their isophotes is nearly elliptical. The shape parameter $a_4$ has 
been
developed in order to quantify the deviations of the isophotes from an 
elliptical
shape and to find correlations between these deviations and other properties of 
the
galaxies. Isophotes whose $a_4$ is negative are called boxy, since their shape
resembles a rectangle, and those with a positive $a_4$ are called disky, because
they are more pointed, lemon-like, than the corresponding ellipse. It is 
generally
believed that a galaxy with disky isophotes has a disk component; however, only 
disks
seen nearly edge-on can be identified by $a_4$ (Carter 1987, Rix \&
White 1990). Moreover, there is a relation between $a_4$ and the radio and X-ray
emission (at least for giant ellipticals), in the sense that boxy ellipticals 
tend
to be the stronger sources. These correlations are shown in the study of Bender 
et
al. (1989), who also find that apparently more flattened 
galaxies are either disky or boxy, whereas
rounder objects tend to have $a_4 \sim 0$ (see their Fig.~1). A similar trend 
for
{\em dwarf}\/ ellipticals as well is shown by Ryden et al. (1999). 
Interestingly,
however, these authors find about a
dozen rather flattened galaxies which do not have boxy or disky isophotes, 
i.e.~strongly flattened 
dwarfs
do not show a gap in the distribution of $a_4$ between boxy and disky. We find 
the
\begin{figure}[t]
\begin{center}
\epsfig{file=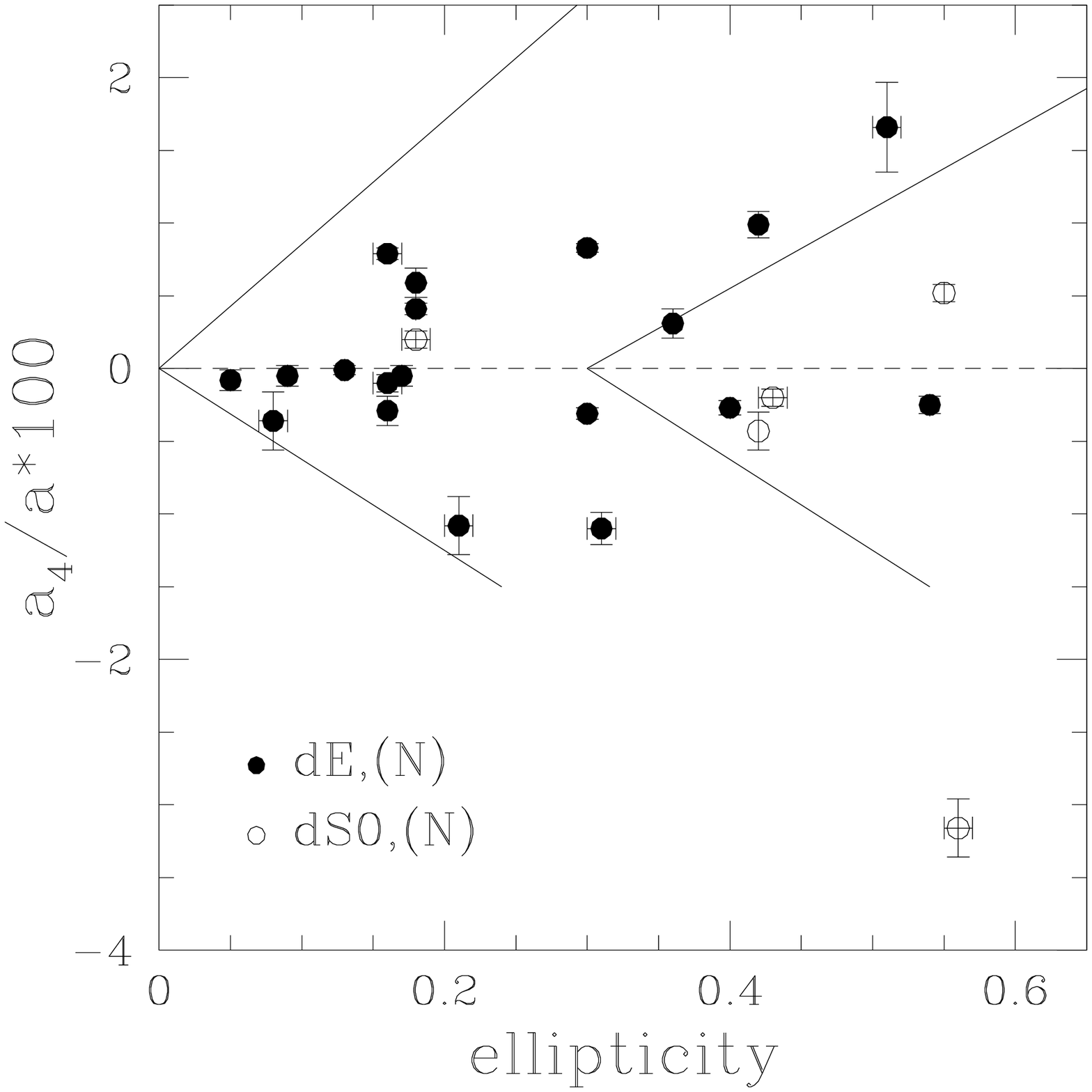,height=80mm,width=80mm}
\end{center}
\caption[]{Ellipticity versus shape parameter represented as $a_4/a*100$ for 
dE,(N)s
(solid circles) and dS0,(N)s (open circles). The lines are the same like in 
Fig.~1
of Bender et al. (1989) and indicate the distribution of giant ellipticals. 
Shown are
also the $1 \sigma$ errors of the averaging.}
\end{figure}
same behavior in our sample. In Fig.~6 we plot ellipticity versus $a_4/a*100$. 
The
lines are the same as in Bender et al. (1989) and bracket the distribution of
giant ellipticals. The general trend that rounder galaxies have $a_4 \sim 0$ 
seems
to be valid for giants as well as for dwarfs. On the other hand, we find 
flattened 
dwarfs
($\epsilon > 0.3$) with $a_4 \sim 0$ -- in contrast to the findings for giants 
and 
in
agreement with Ryden et al. (1999).

However, in view of the profiles for $a_4$ shown in Fig.~1, the significance of 
its
weighted average is not so evident. Most of the profiles show large scattering 
and
frequently change between the boxy and disky regime. Others have a boxy inner
part and a disky outer part (or vice versa), rendering it difficult to asses a 
unique
interpretation to the whole galaxy. For instance, VCC1093 is strongly disky 
in the
central parts, but has nevertheless a negative global $\langle a_4 \rangle$. 
O.~Lehmann (diploma thesis, Basel 2002, unpublished) 
has tried to reproduce the $a_4$ {\em profiles}\/ of a number of individual 
cases
with multi-component models. By evaluating a best set of component parameters a
satisfactory solution could indeed be found for most dwarfs. However, these
models are probably not unique, nor do they seem to be physically very 
meaningful, 
as Stiavelli et al.~(1991) have shown that the isophotal shape of being
either boxy or disky might depend on the viewing angle.
A more promising approach would involve 3D models along the lines of Ryden 
(1992).

Nevertheless, for some individual cases the (projected) shape
parameter is quite straightforward to interpret. Consider for example the 
$a_4$-profile
of IC3388 (Fig.~1): all isophotes in the radius range used are disky, hence, 
it is
justified to classify the galaxy as a whole as disky, suggesting that it hosts a 
disk
component. Or NGC4431, where a hidden bar has been discovered (Barazza et 
al.~2002): the boxyness of the
$a_4$-profile in the central parts clearly reflects the bar and the radius, 
where the
isophotes become disky indicates the extend of the bar.

\subsection{Off-center nuclei}
Even though the position of a nucleus with respect to the overall light 
distribution
might not be considered as part of an isophotal analysis, we discuss this 
phenomenon
in this context, since the measurement of the nuclear offset goes together with 
the
determination of the isophotal parameters (see above). Moreover, offsets found 
not
only represent off-centered nuclei but might also indicate that consecutive
isophotes are not concentric.

In general, the nature of dE nuclei is still unknown; mostly they are regarded 
as
massive compact star clusters which form separate dynamical entities, without 
being
totally decoupled from the rest of the galaxy. They might have formed in the 
last
burst of star formation in the evolutionary transition from dwarf irregulars to 
dEs
(Davies \& Phillipps 1988). Simulations suggest that such nuclei
oscillate about the center of the galaxy (Miller \& Smith 1992, Taga \& Iye 
1998).
However, in these simulations the rotation of the main body is a crucial 
condition, which
would not be complied by most of the dwarf ellipticals. On the other hand, 
Sweatman (1993) showed
in his models that oscillations of central objects can also be explained by 
inherent
motions due to statistical fluctuations, without going back to rotation.

More recently it was suggested that the nuclei could be the result
of the merging of several globular clusters which sunk to the center of the 
galaxy
through dynamical friction (Lotz et al. 2001). In the same study it was also 
shown that
brighter nuclei tend to be in brighter host galaxies. However, a lower globular
cluster specific frequency ($S_N$) suggested for dE,Ns cannot be found; on the 
contrary,
dE,N have a higher $S_N$ than dEs (Miller et al. 1998). Oh \& Lin (2000) studied 
a
similar scenario including extra galactic tidal perturbations accounting for the 
fact that
nucleated dEs are more concentrated to the center of the cluster than 
non-nucleated dEs.
In addition, they found that the nuclei may be slightly off-center within $\sim 
1$ Gyr
after each globular cluster merger event. Indeed, in their study of 78 nucleated 
dwarf
ellipticals and dwarf S0s, Binggeli et al. 2000 (hereafter BBJ) found that $\sim
20\%$ of the objects have off-centered nuclei. They also found a weak 
correlation
between the strength of the offset and the effective surface brightness: fainter 
objects
tend to have larger offsets. However, since the resolution of their data was 
rather low,
the significance of this effect is not clear.
\begin{figure}[t]
\begin{center}
\epsfig{file=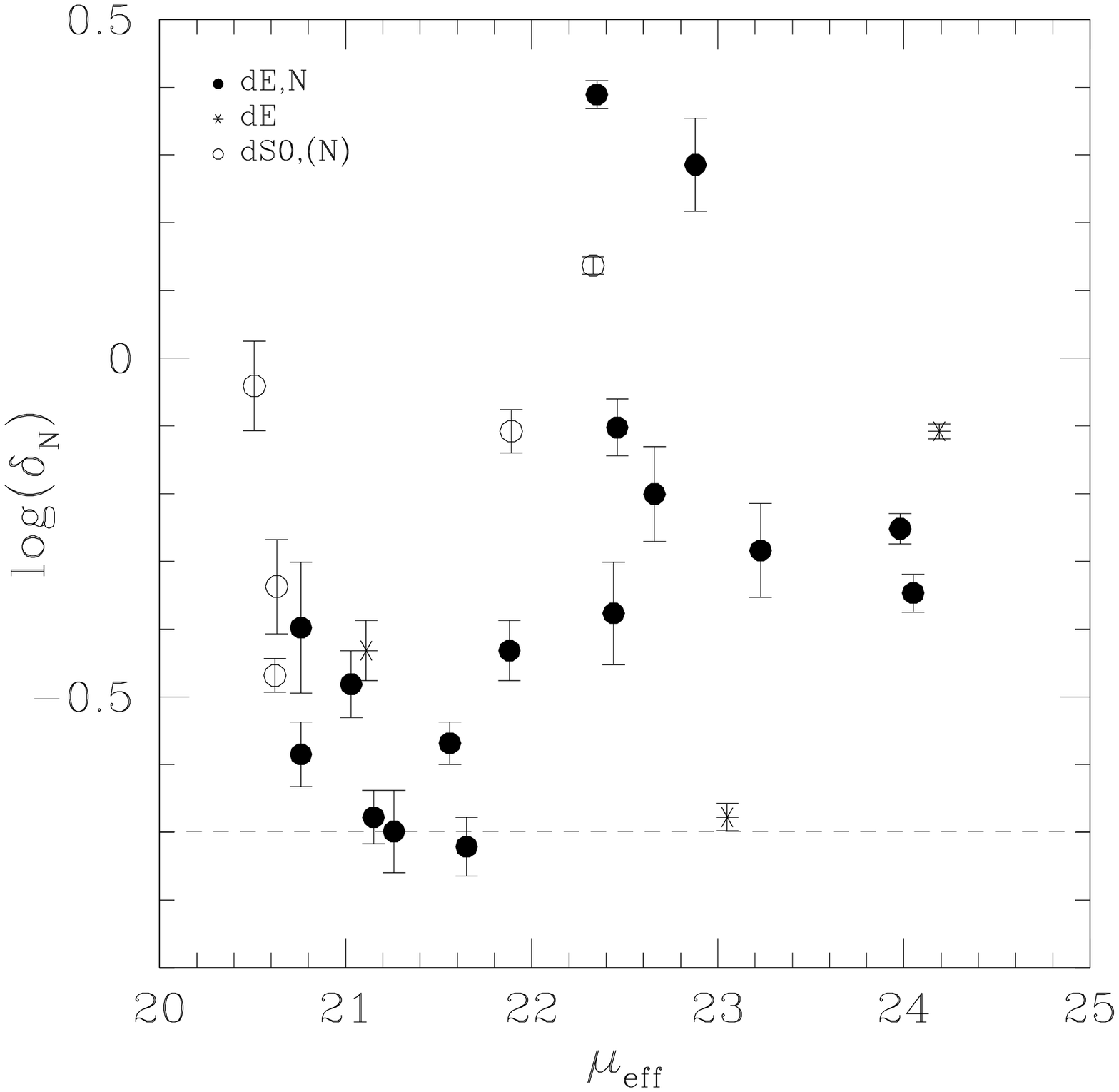,height=80mm,width=80mm}
\end{center}
\caption[]{The logarithm of the nuclear offset versus the effective surface 
brightness
in $R$. The dashed line indicates our lower detection limit of $0.2 \arcsec$, 
i.e.
offsets above this line are real and, in fact, lower limits of the true 
displacements.
The error bars are the $1 \sigma$ errors of the averaging. The meaning of the 
symbols
is indicated.}
\end{figure}

We therefore did the same analysis with our data, i.e. for each galaxy we 
determined
the distances between the centers of the isophotes and the position of the 
nucleus.
In case of an object without nucleus, i.e.~for a dE, the innermost isophote was 
taken as reference.
Due to the photon noise a small offset is always measured, even in the case of 
an
exactly centered nucleus. To assess the resulting systematic and random errors 
we followed BBJ and
performed Monte Carlo
simulations for the method described, using model galaxies with nuclei placed at 
the
center or slighly offset. It turned out that the noise causes a minimum offset 
of
$0 \farcs 2$, i.e. one pixel length, and that a real offset is always 
underestimated,
i.e. offsets measured larger than $0 \farcs 2$ are real and are at the same time
lower limits of the real displacements.
(For details of the measurement and of the error estimation see BBJ). In Fig.~7 
we
plot the logarithm of the nuclear offsets obtained versus effective surface 
brightness
in $R$. Here we distinguish between dEs and dE,N. The dashed line indicates our 
lower
detection limit of $0 \farcs 2$. The relation for the dE,Ns is evident: {\it 
objects with
lower effective surface brightnesses tend to have larger nuclear offsets}, 
confirming the results of BBJ. 
With a correlation coefficient of 0.498 for the 16 dE,Ns, the relation is
significant just at the 95\% level.
This cannot be due to larger errors in determining the 
position of a
nucleus of a fainter object, as these uncertainties always lead to an
underestimation
of the offsets (see above). We would therefore rather assume that, for example, 
the three
dE,Ns with $\mu_{eff} > 23 \,{\rm mag}$ have larger offsets than the ones 
measured.
\begin{figure}[t]
\begin{center}
\epsfig{file=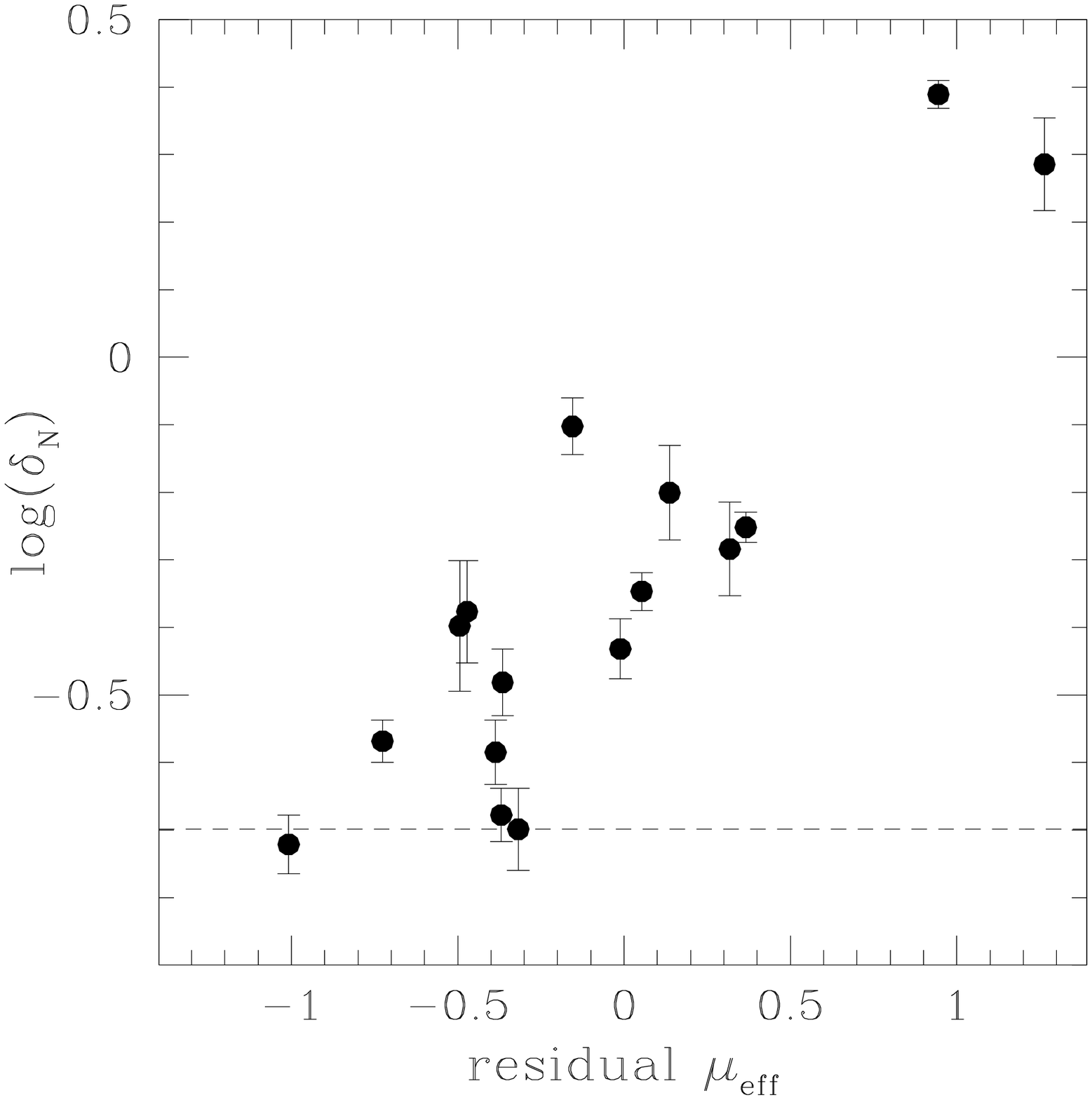,height=80mm,width=80mm}
\end{center}
\caption[]{ The logarithm of the nuclear offset versus the {\em residual}\/
effective surface 
brightness, calculated with respect to the mean relation between 
effective surface brightness
and absolute magnitude (in $R$), for dE,N galaxies only. 
Dashed line as in Fig.~7.
}
\end{figure}

We also plot, in Fig.~8,
the nuclear offset versus the {\em deviation}\/ of the observed $\mu_{eff}$
from the mean $\mu_{eff}$ expected for the observed luminosity.
The underlying mean relation between absolute magnitude and effective surface 
brightness for the 16 dE,N galaxies, to which the residual $\mu_{eff}$ values refer,
is given by the equation $\mu_{eff}$ = 0.545 $M_R$ + 31.39. The relation between
the nuclear offset and the {\em residual}\/ surface brightness (Fig.~8) is 
indeed stronger than the direct relation (Fig.~7). In particular, the two dE,Ns with 
the highest nuclear offsets are not the ones of lowest surface brightness (see Fig.~7),
but they clearly have unusually low surface brightness {\em for their luminosity}\/
(Fig.~8). The correlation coefficient for the relation shown in Fig.~8 is 0.879,
which corresponds to a significance level above 99.9\%.
%An even stronger relation is found when plotting $\log(\delta_N)$ versus the 
%logarithm of
%the effective radius (Fig.~8). In this case, actually, the dEs and the 
%dS0,(N)s
%follow the same relation like the dE,Ns, except six objects forming almost a 
%distinct
%group within a narrow range of effective radii. This outcome seems to be 
%significant,
%since the same partition is seen when using the larger sample of BBJ, which we 
%reexamined
%for this purpose (not shown here).

In the context of the scenarios mentioned above, the relations found in Figs. 
7 and 8
might be interpreted as follows: assuming that a central star cluster would 
oscillate
around the center of the galaxy, we would expect that the oscillations are the 
stronger
the shallower the potential well, i.e.~the fainter the 
effective surface brightness is -- either in an absolute sense (Fig.~7), or 
more likely in
a relative sense with respect to the mean value expected for the luminosity 
(Fig.~8).
On the other hand, if the nucleus is
the result of the merging of several globular clusters, a correlation between 
offset
and $r_{eff}$ or $\mu_{eff}$ is not a natural outcome, e.g. Oh \& Lin (2000) 
do not find a magnification of the effective radius after a merger event in 
their models. The
case of simple oscillations is therefore more likely.

\subsection{Isophotal twists}
In giant elliptical galaxies isophotal twists are a quite common phenomenon
(Jedrzejewski 1987, Kormendy \& Djorgovski 1989), usually divided into outer and 
inner
twists. The latter are believed to be mainly caused by tidal effects (Kormendy 
1982) or
even by errors in the flat-fielding and background subtraction process (Fasano 
\&
Bonoli 1990), while inner twists are attributed to the two-component nature of 
many
ellipticals (Nieto et al. 1992). Indeed, there are similar features 
in the
profiles of isophotal parameters of Es and SB0s galaxies. Moreover, there is a 
correlation
between isophotal twists and flattening in ellipticals (Galletta 1980). Finally, 
Ryden
et al. (1999) determined isophotal twists in their sample of dwarf ellipticals, 
finding
however only a weak relation with luminosity: brighter galaxies tend to have 
smaller
values of isophotal twists.

Using the usual range ($4 \arcsec < r,\mu \leq 26{\rm mag}/\sq\arcsec$) we
determined the largest isophotal twist, $\Delta P.A.$, of the galaxies in our 
sample
(column 10 of Table 3). In Fig.~9 we plot the twists derived versus 
ellipticity.
\begin{figure}[t]
\begin{center}
\epsfig{file=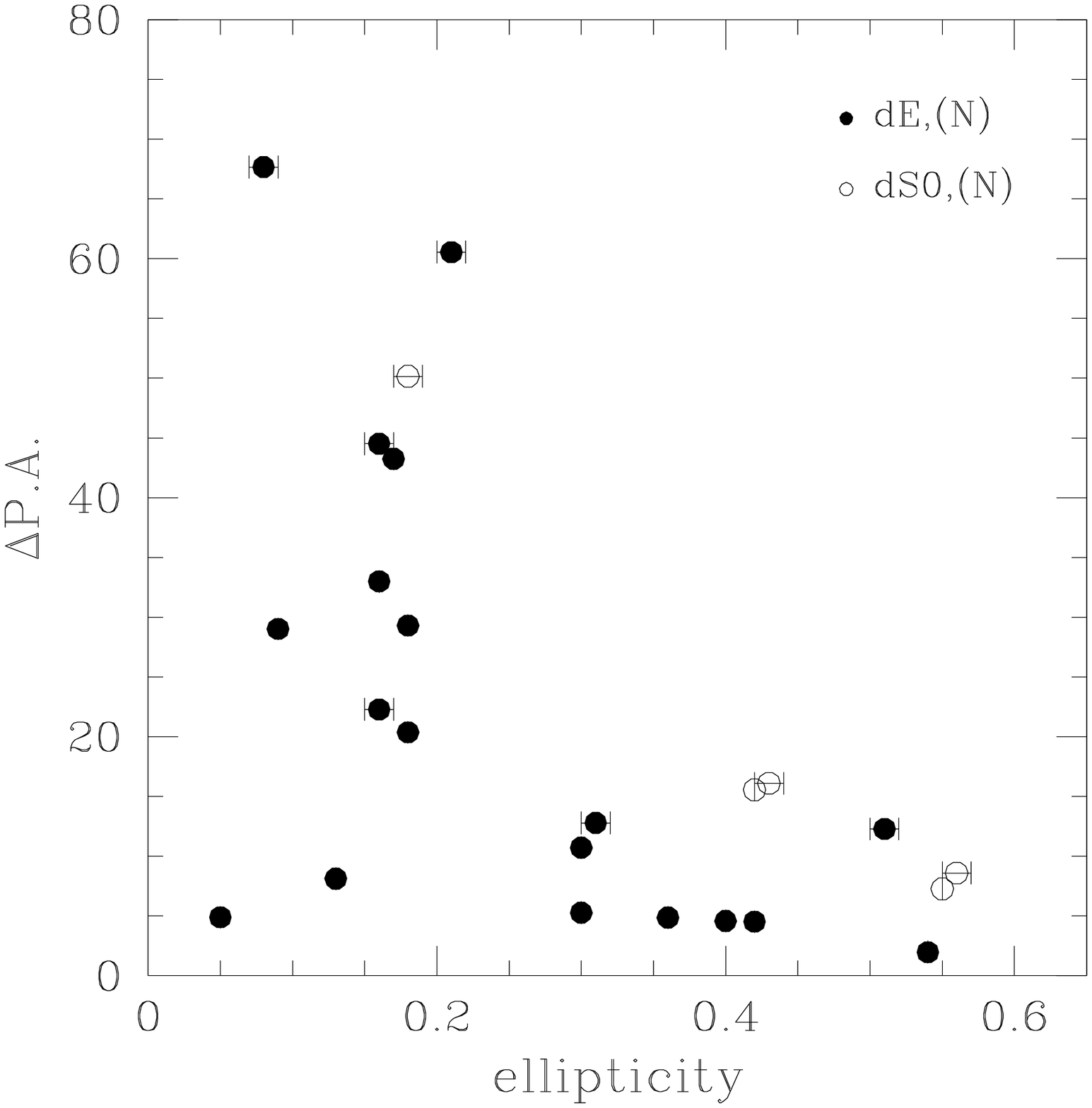,height=80mm,width=80mm}
\end{center}
\caption[]{Isophotal twist, $\Delta P.A.$, versus ellipticity. The meaning of 
the
symbols is indicated.}
\end{figure}
It is striking that {\it strong isophotal twists are only exhibited by galaxies 
with
$\epsilon < 0.3$}; more flattened
objects show only weak twists. The relation found is very similar to the one 
presented
by Galletta (1980) for giant ellipticals (see his Fig.~1). However, it could 
be
argued that ellipticities and, above all, position angles of round galaxies have 
in
general larger errors. Since for a round isophote already a small change of its 
shape,
probably caused by noise, can lead to a large change of the position angle, we 
might
expect a strong scatter in the radial profile of the galaxy and thus a larger 
twist.
Nevertheless, we believe that the relation found is real for the following 
reasons:
(1) We measure inner twists, as the typical position where the twists occur 
is at
$\sim 1.31$ effective radii (average for galaxies with $\Delta P.A. > 20$).
Measurement errors in these bright regions of the galaxies are negligible. 
Moreover,
the outer limit considered is at a surface brightness level of $26$ mag, which 
is well
within the confidence level
(see Table 2). The errors of position angle measurements in this range are 
indeed
very small (see error bars of the $P.A.$ profiles in Fig.~1), hence large
scatter caused by noise are very unlikely. (2) The border line between galaxies 
with and
without strong twists occurs at an ellipticity of $\sim 0.3$, which is still 
fairly round,
i.e. there is no reason why there should not also be strongly twisted dwarfs
with $\epsilon$ slightly larger than $0.3$. (3) Even if
we would assume that the determination of twists for round galaxies is due to
photometric inaccuracy,
the spread of values exhibited by these objects must be real, as it does not 
depend
on luminosity (no correlation was found between twist and total magnitude). (4)
Considering the $P.A.$ profiles in Fig.~1, it is evident that the variations 
of the
position angles are systematic and not irregular, as would be expected, if they
were caused by measurement errors. Strong oscillations are only found in 
galaxies
for which the presence of a spiral structure has been shown (IC0783, IC3328) 
and,
hence, the strong scatter is explicable.

\begin{figure}[t]
\begin{center}
\epsfig{file=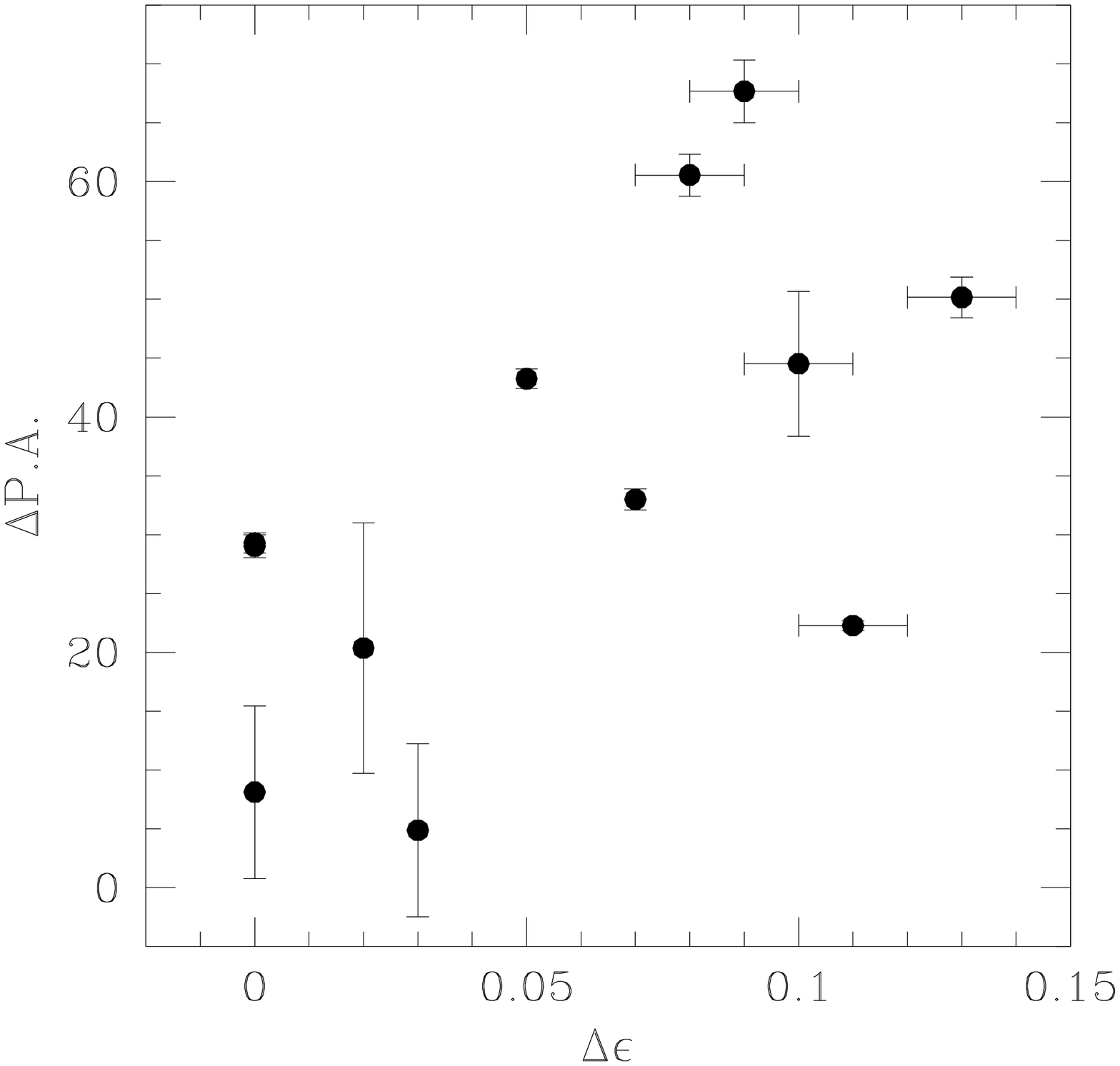,height=80mm,width=80mm}
\end{center}
\caption[]{Isophotal twist, $\Delta P.A.$, versus the corresponding change of
ellipticity, $\Delta \epsilon$, for galaxies with $\epsilon < 0.3$. (Two plot 
symbols
strongly overlap at $\Delta \epsilon \approx 0$, $\Delta P.A. \approx 29$.)}
\end{figure}
In order to asses the meaning of 
the twists observed it is important to consider the fact that we
can only measure apparent ellipticities. As the distribution of intrinsic
ellipticities for dwarf ellipticals has a maximum at $\sim 0.3$ with a sharp 
decline
for rounder shapes (Binggeli \& Popescu 1995), most of the objects with 
$\epsilon
< 0.3$ in our sample might suffer from projection effects, i.e. they appear 
rounder
than they really are. On the other hand, assuming that these objects are 
triaxial
ellipsoids, we would expect to observe an apparent isophotal twist, if the 
intrinsic
axial ratios would change with radius
(Binney 1978). Indeed, large twists are only exhibited by galaxies with 
$\epsilon <
0.3$ and might therefore be due to {\it projection} and not to intrinsic changes 
of the
position angles. In Figure 10 the isophotal twists are plotted versus the change 
of
ellipticity, measured in the same radial range in which the twist is observed. 
Only
galaxies with $\epsilon < 0.3$ are shown. A rather striking correlation is in 
fact
evident: {\it strong twists seem to occur only when a considerable change of the
ellipticity takes place as well}\/ (the correlation coefficient for the 12 galaxies
plotted in Fig.~10 is 0.600, corresponding to a level of significance slightly 
above 95\%). This strongly supports the conjecture that
the twist effect (Fig. 9) is due to the {\em projection of intrinsically 
triaxial
ellipsoids}.

Besides the twist-ellipticity relation, we also found a weaker correlation 
between
twist and effective surface brightness. Thus, a connection between isophotal 
twist
and nuclear offset might be expected and is indeed present, at least for the 
dE,Ns.
Fig.~11 shows that {\it galaxies with stronger twists have larger nuclear 
offsets}\/ (correlation coefficient for the 16 dE,Ns = 0.506, giving a
statistical significance better than 95\%).
As the measurement of the offsets should not be affected by projection 
effects,
the origin of the relation in Figure 11 is difficult to understand. However,
assuming that the isophotal twists are real, the connections between effective
surface brightness/radius, offset, and twist might indicate that less compact 
objects also
tend to have stronger anomalies in their isophotal properties.
\begin{figure}[t]
\begin{center}
\epsfig{file=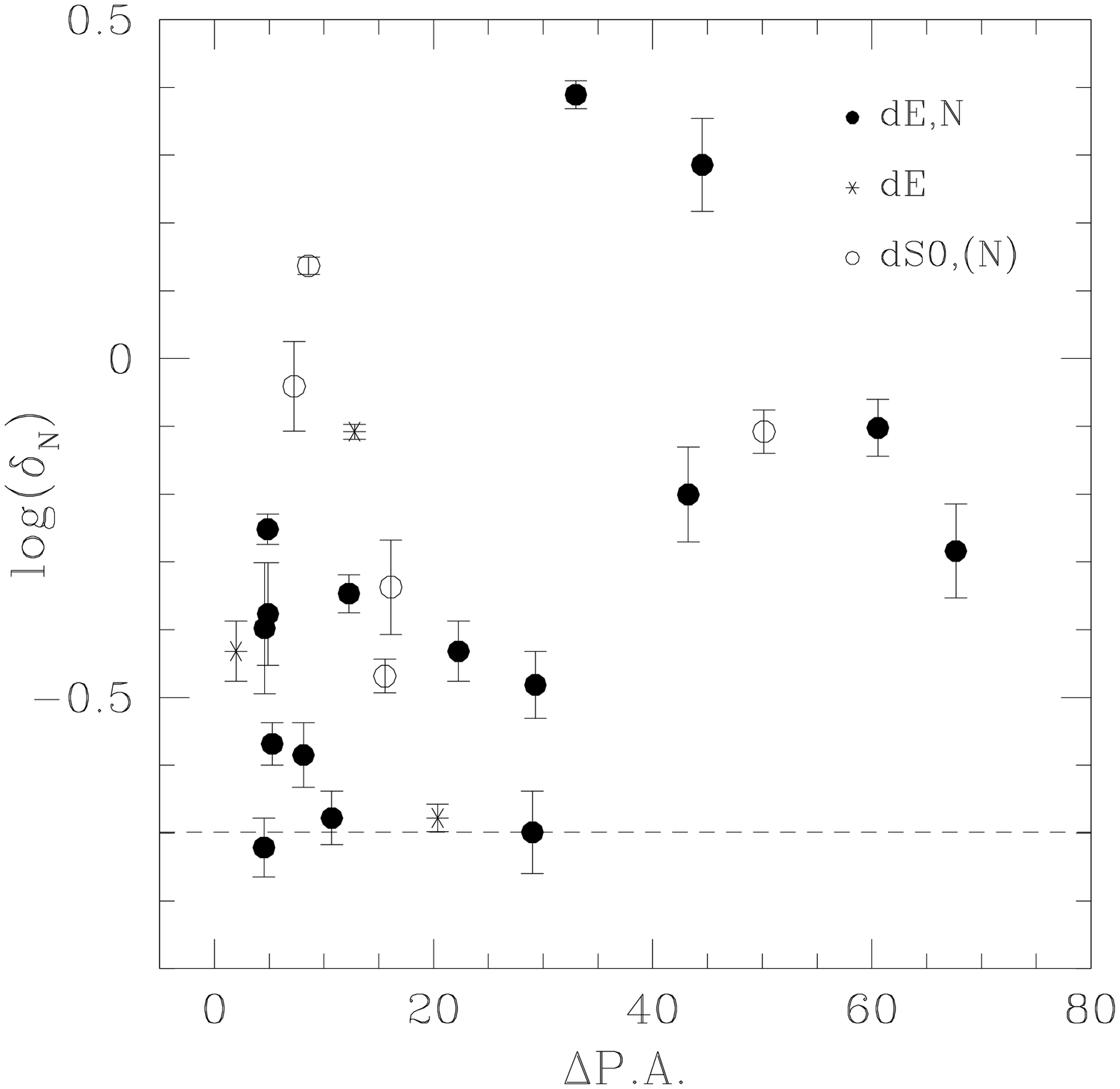,height=80mm,width=80mm}
\end{center}
\caption[]{Isophotal twist, $\Delta P.A.$, versus the logarithm of $\delta_N$. 
The
dashed line indicates our lower detection limit of $0 \farcs 2$. The meaning of 
the
symbols is indicated.}
\end{figure}

\subsection{Ellipticity profiles}
Most of the galaxies show a rather steep
central rise in the ellipticity profile, particularly obvious in the brightest 
nucleated
objects (see Fig.~1). 
This is not the nucleus itself, which is much smaller in size, but it might be 
due to the gravitational influence of the nucleus on the orbits of the nearby 
stars. Indeed, simulations indicate 
that
already a rather low massive object in the center can scatter stars off the 
central
density cusp and therefore disrupt box orbits and forcing the stars on circular 
paths
(Gerhard \& Binney 1985). If the nucleus is more massive, probably hosting a 
black
hole, it can even shape the whole galaxy, leading to a rounder object. This was 
in fact
predicted by Norman (1986) and confirmed through observations by BBJ (see their
Appendix).

\section{Summary and Conclusions}
\subsection{Surface photometry}
We have carried out surface photometry (in $B$ and $R$) and a detailed isophotal
analysis (in $R$) for 25 early-type dwarf galaxies in the Virgo cluster. The 
sample
considered mainly consists of nucleated dwarf ellipticals and has a mean 
absolute
magnitude in $B$ of $\langle M_{B_T} \rangle = -15.76$ and a mean colour of
$\langle B-R \rangle = 1.27$.

We provide surface brightness profiles in both filters and radial colour 
gradients.
All profiles have been fitted by a S\'ersic model and the corresponding 
parameters
have been derived. Plotting these best fitting parameters together with the 
ones for giant
ellipticals and local group dwarfs versus absolute magnitude (Fig.~2), we 
confirm
the finding of Jerjen \& Binggeli (1997) and Jerjen, Binggeli \& Freeman 
(2000) that the cluster dwarfs perfectly bridge the
gap between giant ellipticals and dwarf spheroidals, indicating that there is 
{\em one, continuous family}\/ of spheroidal stellar systems, provided
the central few 100 pc of these galaxies is not considered. 
However, considering in detail the fits obtained, we find
several profiles with rather large deviations from the models, in particular in 
the more central parts. By defining a mean,
absolute residual, $\langle S_{res} \rangle$, taking into account also the 
errors 
of the profiles, we find that bright galaxies, 
having $\mu_{eff} < 22$ mag, show the strongest deviations from the S\'ersic 
form. 
We suggest that in addition to the objects
with disk structure (spiral or bar) signatures (Jerjen, Kalnajs \& Binggeli 
2000,
Barazza et al. 2002) where strong residuals might be expected, some of 
the bright dwarfs may be quite complex in structure, possibly being 
two-component 
systems as well.

\subsection{Isophotal analysis}
By fitting ellipses to the isophotes, we have derived radial profiles for the
ellipticity, position angle, major axis, and shape parameter $a_4$
(Fig.~1). For these parameters, as well as for the offsets of the nuclei we have
determined mean (global), luminosity-weighted values (Table 3).
Plotting ellipticity versus $a_4/a*100$ we find a similar distribution as Bender
et al. (1989) for giant ellipticals. However, as in the dE sample of 
Ryden et al. (1999) and in contrast to giant Es,
there is no gap in the distribution between disky and boxy for highly flattened
galaxies, i.e.~there are highly flattened dEs that are perfectly elliptical. 
The meaning of a {\em mean}\/ $a_4$-value is not so clear, however, as
most of the galaxies have a $a_4$-profile that changes sign (from boxy to disky, 
or
vice versa) along the galactocentric radius. Almost any behaviour of the
$a_4$ profile can easily be reproduced with
multi-component (many parameter) models, but in lack of any knowledge about the 
inclination(s),
i.e.~the {\em intrinsic}\/ shapes of the components, it seems impossible to 
construct {\em unique}\/ models for the isophotal structure of these galaxies.

As most of the sample galaxies are nucleated, we have also searched for the 
presence
of off-center nuclei. Most nuclei are indeed slightly off center.
Defining $\delta_N$ as the offset of the nucleus 
from the center of the overall light distribution, we have looked for 
systematic relations between 
nuclear offsets
and other galaxy properties. We confirm the existence of a relation between 
$\delta_N$ and
the effective surface brightness, which was first suggested by Binggeli et 
al.~(2000): galaxies with fainter effective
surface brightnesses tend to have larger nuclear offsets. 
An even stronger relation is found to hold between $\delta_N$ and the 
{\em residual}\/ effective radius, determined with respect to the mean relation between
effective surface brightness and absolute magnitude, i.e.~dwarfs with lower than
average surface brightness {\em at a given magnitude}\/ have larger nuclear offsets.
The most plausible explanation of the phenomenon is that the nuclei
are simply oscillating about the centers in the shallow potential wells of 
these galaxies, 
as suggested by numerical simulations (Miller \& Smith 1992, Taga \& Iye 1998).
The shallower the potential well (the lower the surface brightness), the higher 
the expected amplitude of the oscillations. 

In addition, we determined isophotal twists, i.e. the largest changes of the
position angles of the major axes within the range used for the isophotal 
analysis.
We find a clear dependence of the twists measured on ellipticity in that
large
twists occur only in galaxies with $\epsilon < 0.3$. On the other hand, most of 
these
galaxies are likely intrinsically more flattened, as the distribution of 
intrinsic
ellipticities has a maximum at $\sim 0.3$. 
Assuming that the objects are intrinsically triaxial ellipsoids, 
we suggest that the twists observed, always being accompanied by changing 
axis ratios, are caused by projection effects.

In view of the results of the isophotal analysis we conclude that 
{\em less compact
dwarf galaxies generally 
tend to have stronger irregularities}\/ like off-centered nuclei or
twisted isophotes, than more compact systems. However, with the available data 
we cannot decide whether 
these
properties are caused by external perturbations or are due to the presence of
substructures in these objects. Moreover, projection effects may play a crucial
role in producing certain isophotal properties. Assuming that dwarf elliptical 
galaxies
are intrinsically triaxial systems, the twists observed could be explained in 
terms
of projection effects caused by changing axial ratios in apparently round 
galaxies
that are intrinsically more flattened.

\begin{acknowledgements}
We thank the referee, Dr.~N.~Caldwell, for his constructive comments.
F.D.B and B.B. are grateful to the Swiss National Science Foundation for
financial support.
\end{acknowledgements}


\begin{thebibliography}{}

\bibitem{}
Barazza, F.D., Binggeli, B., Jerjen, H. 2002, A\&A, 391, 823

\bibitem{}
Bender, R., M\"ollenhoff, C. 1987, A\&A, 177, 71

\bibitem{}
Bender, R., Surma, P., D\"obereiner, S., M\"ollenhoff, C., Medejsky, R. 1989, 
A\&A,
217, 35

\bibitem{}
Binggeli, B., Cameron, L.M. 1991, A\&A, 252, 27

\bibitem{}
Binggeli, B., Cameron, L.M. 1993, A\&AS, 98, 297

\bibitem{}
Binggeli, B., Jerjen, H. 1998, A\&A, 333, 17

\bibitem{}
Binggeli, B., Popescu, C.C. 1995, A\&A, 298, 63

\bibitem{}
Binggeli, B., Popescu, C.C., Tammann, G.A. 1993, A\&AS, 98, 275

\bibitem{}
Binggeli, B., Sandage, A., Tarenghi, M. 1984, AJ, 89, 64

\bibitem{}
Binggeli, B., Sandage, A., Tammann, G.A. 1985, AJ, 90, 1681 (VCC)

\bibitem{}
Binggeli, B., Popescu, C.C., Tammann, G.A. 1993, A\&AS, 98, 275

\bibitem{}
Binggeli, B., Barazza, F.D., Jerjen, H. 2000, A\&A, 359, 447 (BBJ)

\bibitem{}
Binney, J. 1978, MNRAS, 183, 779

\bibitem{}
Bothun, G.D., Mould, J.R., Caldwell, N., Mac\,Gillivray, H.T. 1986, AJ, 92, 
1007

\bibitem{}
Byun, Y.-I., Grillmair, C.J., Faber, S.M., Ajhar, E.A., Dressler, A.,
Kormendy, J., Lauer, T.R., Richstone, D., Tremaine, S. 1996, AJ, 111, 1889

\bibitem{}
Caldwell, N. 1983, AJ, 88, 804

\bibitem{}
Caldwell, N., Bothun, G.D. 1987, AJ, 94, 1116

\bibitem{}
Caon, N., Capaccioli, M., D'Onofrio, M. 1993, MNRAS, 265, 1013

\bibitem{}
Carter, D. 1987, ApJ, 312, 514

\bibitem{}
Conselice, C.J., Gallagher III, J.S., Wyse, R.F.G. 2001, ApJ, 559, 791

\bibitem{}
De Rijcke, S., Dejonghe, H., Zeilinger, W.W., Hau, G.K.T. 2001, ApJ, 559, 21

\bibitem{}
De Vaucouleurs, G. 1959, Handbuch der Physik 53, ed. S. Fl\"ugge (Springer,
Berlin), 275

\bibitem{}
Davies, J.I., Phillipps, S. 1988, MNRAS, 233, 553

\bibitem{}
Durrell, P.R. 1997, AJ, 113, 531

\bibitem{}
Faber, S.M., Tremaine, S., Ajhar, E.A., Byun, Y.-I., Dressler, A.,
Gebhardt, K., Grillmair, C., Kormendy, J., Lauer, T.R., Richstone, D. 1997, 
AJ, 114, 1771

\bibitem{}
Fasano, G., Bonoli, C. 1990, A\&A, 234, 89

\bibitem{}
Ferguson, H.C. 1989, AJ, 98, 367 

\bibitem{}
Ferguson, H.C., Binggeli, B. 1994, A\&AR, 6, 67

\bibitem{}
Ferguson, H.C., Sandage, A. 1988, AJ, 96, 1520

\bibitem{}
Galletta, G. 1980, A\&A, 81, 179

\bibitem{}
Geha,M., Guhathakurta, P., van der Marel, R. 2002, AJ, 124, 3073

\bibitem{}
Gerhard, O.E., Binney, J. 1985, MNRAS, 216, 467

\bibitem{}
Harris, W.E., Durrell, P.R., Pierce, M.J., Secker, J. 1998, Nature, 395, 45

\bibitem{}
Hoffman, G.L., Williams, H.L., Salpeter, E.E., Sandage, A., Binggeli, B. 1989, 
ApJS,
71, 701

\bibitem{}
Jedrzejewski, R.I. 1987, MNRAS, 226, 747

\bibitem{}
Jerjen, H., Binggeli, B. 1997, in The Nature of Elliptical Galaxies, 
2nd Stromlo
Symposium, ed.~M.~Arnaboldi et al., ASP Conf.~Ser.~Vol.~116, p.~239

\bibitem{}
Jerjen, H., Binggeli, B., Freeman, K.C. 2000, AJ, 119, 593

\bibitem{}
Jerjen, H., Kalnajs, A., Binggeli, B. 2000, A\&A, 358, 845

\bibitem{}
Jerjen, H., Binggeli, B., Barazza, F.D. 2003, in preperation

\bibitem{}
Jerjen, H., Kalnajs, A., Binggeli, B. 2001, in Galaxy Disks and Disk Galaxies,
ASP Conf. Ser., Vol. 230, eds. J.G. Funes S.J. \& E.M. Corsini, Astronomical 
Society
of the Pacific, p.239

\bibitem{}
King, I.R. 1966, AJ, 71, 64

\bibitem{}
Kormendy, J. 1982, in Morphology and Dynamics of
Galaxies, Martinet, L., Mayor, M. (eds.),
12th Advanced Course of the Swiss Society of Astronomy and 
Astrophysics. Sauverny, Geneva Observatory

\bibitem{}
Kormendy, J., Djorgovski, S. 1989, ARA\&A, 27, 235

\bibitem{}
Landolt, A.U. 1992, AJ, 104, 340

\bibitem{}
Lauer, T.R., Ajhar, E.A., Byun, Y.-I., Dressler, A., Faber, S.M., Grillmair, C.,
Kormendy, J., Richstone, D., Tremaine, S. 1995, AJ, 110, 2622

\bibitem{}
Lotz, J.M., Telford, R., Ferguson, H.C., Miller, B.W., Stiavelli, M., Mack, J. 
2001
ApJ, 552, 572

\bibitem{}
Miller, B.W., Lotz, J.M., Ferguson, H.C., Stiavelli, M., Whitmore, B.C. 1998, 
ApJ,
508, L133

\bibitem{}
Miller, R.H., Smith, B.F. 1992, ApJ, 393, 508

\bibitem{}
Nieto, J.-L., Bender, R., Poulain, P., Surma, P. 1992, A\&A, 257, 97

\bibitem{}
Norman, C. 1986, in Star-Forming Dwarf
Galaxies and related objects, Kunth, D., Thuan, T.X., Van, J.T.T (eds.),
Gif sur Yvette: Edition Frontieres p. 477

\bibitem{}
Oh, K.S., Lin, D.N.C. 2000, ApJ, 543, 620

\bibitem{}
Rix, H.-W., White, S.D.M. 1990, ApJ, 362, 52

\bibitem{}
Ryden, B.S. 1992, ApJ, 386, 42

\bibitem{}
Ryden, B.S., Terndrup, D.M., Pogge, R.W., Lauer, T.R. 1999, ApJ, 517, 650

\bibitem{}
Sandage, A., Binggeli, B. 1984, AJ, 89, 919

\bibitem{}
Sandage, A., Binggeli, B., Tammann, G.A. 1985, AJ, 90, 1759

\bibitem{}
Schlegel, D.J., Finkbeiner, D.P., Davis, M. 1998, ApJ, 500, 525

\bibitem{}
S\'ersic, J.-L. 1968, Atlas de galaxias australes, Observatorio Astronomico,
Cordoba

\bibitem{}
Simien, F., Prugniel, Ph. 2002, A\&A, 384, 371

\bibitem{}
Stiavelli, M., Londrillo, P., Messina, A. 1991, MNRAS, 251, 57

\bibitem{}
Stiavelli, M., Miller, B.W., Ferguson, H.C., Mack, J., Whitmore, B.C., Lotz, 
J.M. 2001, AJ, 121, 1385

\bibitem{}
Taga, M., Iye, M. 1998, MNRAS, 299, 111

\bibitem{}
Thomas, D., Bender, R., Hopp, U., Maraston, C., Greggio, L. 2003, in
The Evolution of Galaxies. III. Kiel Euroconference, G.~Hensler et al. (eds.),
Kluwer

\bibitem{}
Vader, J.P., Vigroux, L., Lachi\`eze-Rey, M., Souviron, J. 1988, A\&A, 203, 217

\bibitem{}
Young, C.K., Currie, M.J. 1994, MNRAS, 268, L11

\end{thebibliography}
\end{document}